\documentclass[letterpaper,onecolumn,accepted=2022-06-24]{quantumarticle}
\pdfoutput=1
\usepackage[utf8]{inputenc}

\usepackage{amsthm,amsfonts,framed}
\usepackage{physics, graphicx, hyperref}
\newcommand{\dbyd}[1]{\frac{\partial}{\partial #1}}

\usepackage{amsmath,amssymb,amsthm}

\def\bra#1{\langle#1|}
\def\ket#1{|#1\rangle}

\newtheorem{thm}{Theorem}
\newtheorem{cor}{Corollary}
\newtheorem*{thm*}{Theorem}

\newtheorem*{conj*}{Conjecture}

\numberwithin{equation}{section}

\DeclareMathOperator{\E}{\mathbb{E}}

\begin{document}
\title{Bounds on approximating Max $k$XOR with quantum and classical local algorithms}
\author[1,2,3,4]{Kunal Marwaha}
\orcid{0000-0001-9084-6971}
\email{kmarw@uchicago.edu}
\author[1,2]{Stuart Hadfield}
\orcid{0000-0002-4607-3921}
\affil[1]{Quantum Artificial Intelligence Laboratory (QuAIL), NASA Ames Research Center, Moffett Field, CA}
\affil[2]{USRA Research Institute for Advanced Computer Science (RIACS), Mountain View, CA}
\affil[3]{Berkeley Center for Quantum Information and Computation, University of California, Berkeley, CA}
\affil[4]{Department of Computer Science, University of Chicago, Chicago, IL}

\maketitle

\begin{abstract}
	We consider the power of
	local algorithms for approximately solving Max $k$XOR, a generalization of two constraint satisfaction problems previously studied with classical and quantum algorithms (MaxCut and Max E3LIN2). 
	In Max $k$XOR each
	constraint is the XOR of exactly $k$ variables and a parity bit. 
	On instances with either \emph{random signs} (parities) or \emph{no overlapping clauses} and $D+1$ clauses per variable, we calculate the 
	expected satisfying fraction of the depth-1 QAOA from 
	\cite{farhi2014quantum} and compare with a generalization of the local threshold algorithm from \cite{hirvonen2017large}. Notably, the quantum algorithm outperforms the threshold algorithm for $k > 4$.
    
    On the other hand, we highlight potential difficulties for the QAOA to achieve computational quantum advantage on this problem.
    We first compute a tight upper bound on the maximum satisfying fraction of nearly all large random regular Max $k$XOR instances by numerically calculating the ground state energy density $P(k)$ of a mean-field $k$-spin glass \cite{sen2017optimization}. The upper bound grows with $k$ much faster than the performance of both one-local algorithms. We also identify a new obstruction result for low-depth quantum circuits (including the QAOA) when $k=3$, generalizing a result of \cite{bravyi2019obstacles} when $k=2$. We conjecture that a similar obstruction exists for all $k$.
\end{abstract}


\section{Introduction}

While
quantum approaches to combinatorial
optimization 
such as the quantum approximate optimization algorithm (QAOA) \cite{farhi2014quantum} have attracted much recent attention, relatively few rigorous results are known concerning their potential for quantum advantage. To help clarify this, a recent line of work has sought to compare the performance of low-depth realizations of QAOA with \textit{local} classical algorithms for specific problems.
This is because QAOA is also \textit{local}; at any fixed depth, QAOA shares the same computational restrictions as a local classical algorithm of the same depth.
In particular, these algorithms assign a value to each variable that depends only on a fixed local neighborhood around that 
variable.\footnote{See for instance \cite{barak2021classical} for precise definitions and further discussion of local algorithms.}
We continue this research direction
for the general class of Max $k$XOR constraint satisfaction problems, for $k\geq 2$.
\newline

A Max $k$XOR instance consists of a set of clauses each acting on a constant number $k$ of the problem variables, where a clause is satisfied if its variables' combined parity is equal to a fixed value of odd or even. 
The goal is to find an assignment maximizing the fraction of satisfied clauses. (We use the fraction of satisfied clauses as the measure of 
performance for the algorithms we consider throughout this work.\footnote{In local algorithms, the number of clauses satisfied on a high-girth graph is the same regardless of whether the graph is bipartite or very far from it. So we find it more natural to focus on the satisfying fraction rather than the ratio of clauses satisfied over optimal number of clauses satisfied.})  For the case of~$n$ variables 
$x_1,...x_n\in\{0,1\}$ 
and $m$ clauses, this can be stated as
\begin{align*}
    \max_{x_1,...,x_n} \, & \sum_{\{\ell_1,...,\ell_k\} \in M}   x_{\ell_1} + ... + x_{\ell_k}  + a_{\ell_1...\ell_k} \mod 2, 
\end{align*}
where $M\subseteq {[n] \choose k}$, $|M|=m$, and the values  $a_{\ell_1...\ell_k}\in\{0,1\}$ specify the parity (sign) of each clause. 
For $k \geq 2$, this problem is NP-hard to approximate above a certain value~\cite{haastad2001some}. The case $k=2$ with all parities
even ($a_{\ell_1...\ell_k}=0$) is the Maximum Cut problem, which has been 
frequently studied in the QAOA literature~\cite{farhi2014quantum,Wang2018,Marwaha2021localclassicalmax,barak2021classical}. 
\newline

Shortly after QAOA was proposed,
\cite{farhi2015quantum} showed that the depth-1 version (i.e., the QAOA circuit with a single alternation of phase and mixing layers) outperformed all known classical algorithms on Max 3XOR.
This result attracted a lot of attention, contributing to the popularity of QAOA and in particular the search for possible NISQ optimization advantage (see discussion in \cite{barak2021classical}). This attention also garnered healthy skepticism: soon after,
a classical algorithm was shown to have  better performance scaling 
on general instances \cite{barakmaxklin2}, although the proof was not optimized for achieving the best possible constant. We note that while deeper QAOA circuits can in principle only improve on the depth-1 performance, their rigorous analysis remains quite challenging. 
\newline

It turns out that the performance of the depth-1 QAOA matches the scaling of the classical algorithm on both the set of instances with \emph{random signs}
(referred to as \emph{typical instances} in \cite{farhi2015quantum, lin2016performance})
and those with \emph{no overlapping clauses} (defined as \emph{no overlapping constraints} in \cite{barakmaxklin2}).
The former means that each clause has parity independently odd or even with the same probability.
The latter, also called \emph{triangle-free} instances, means that any two variables are involved in at most one clause and that they share no neighbors outside of that clause. 
We interchange the above qualifiers throughout this work.
This may hint at a general correspondence between the two settings. A one-local algorithm can't distinguish between the case where the clauses adjacent to a particular clause are independently satisfied because the signs on the clauses are random (the former), or because adjacent variables themselves are distinct (the latter). Depth-1 QAOA always
gives a one-local algorithm on problems where each variable participates in a bounded number of clauses.
\newline

This also suggests that local classical algorithms may perform about as well as 
QAOA on Max $k$XOR, at least at low or constant QAOA depth. 
Candidates for such algorithms include the
randomized 
approach of  \cite{hirvonen2017large} 
and the local tensor algorithm \cite{hastings2019classical}.
For one-local algorithms on instances without overlapping clauses, there is only one subgraph: the chance of one clause being satisfied is exactly the average satisfying fraction on the whole graph.
These instances are thus much simpler to analyze.
However, local algorithms will not achieve the optimal satisfying fraction for all instances when $k \ge 4$ is even, because of the overlap gap property \cite{Chen_2019}. This separation likely exists whenever the overlap gap property holds.\footnote{In fact, a recent result shows that the overlap gap property also obstructs low-depth QAOA on these problems~\cite{chou2021limitations}.} Nonetheless, \cite{alaoui2020optimization,montanari2021optimization} suggest a local approximate message-passing algorithm that is optimal on many problems where the overlap gap property does not hold.
\newline

Low-depth quantum algorithms have other limitations. Their performance is related to the No Low-energy Trivial States (NLTS) conjecture, which posits the existence of families of local Hamiltonians whose low energy states are all nontrivial.  A ``trivial'' state is one that can be achieved starting from a product state $|S\rangle$ and with a low-depth quantum circuit $U$. The NLTS conjecture is true for all quantum circuits $U$ if the quantum PCP conjecture holds \cite{hastings2013trivial}, but it is unknown in full generality. Reference~\cite{bravyi2019obstacles} shows that when $U$ and $|S\rangle$  obey
a $\mathbb{Z}_2$ symmetry, there is a family of local Hamiltonians with the NLTS property. The Hamiltonians they construct are instances of MaxCut (and Max $2$XOR); since the associated unitaries for QAOA have $\mathbb{Z}_2$ symmetry, it follows the QAOA is obstructed from the optimal solution until at least logarithmic-depth. We emphasize however that the performance of QAOA is not well-understood in most cases where its depth is sufficiently large as to 
overcome obstructions due to symmetry or locality (as \cite{farhi2020quantumTypicalCase} calls ``seeing the whole graph'').\footnote{See \cite{ho2018efficient} for examples (outside of combinatorial optimization) of linear-depth QAOA preparing nontrivial ground states.}
\newline

An ideal algorithm on Max $k$XOR finds an assignment with the highest possible satisfying fraction. For random instances, the optimal value has a connection to the theory of spin glasses.\footnote{The connection is only to the optimal satisfying fraction, not the assignment that produces it.}  These systems take exponentially long times to relax into the ground state, which has an energy density related to the free energy at zero temperature. In the simplest model (the Sherrington-Kirkpatrick model), this value was heuristically calculated in the 1970s by Parisi in a breakthrough work which introduced the notion of ``replica symmetry breaking'' \cite{Parisi_1980}. This ``trick'' works by averaging many systems and finding the minimum value of a functional. It took twenty-five years to make this heuristic calculation into a rigorous proof \cite{Talagrand2006ThePF}. Recent techniques find the Parisi constant for many spin glasses directly at zero temperature \cite{auffinger2016parisi}, and connections have been drawn from Parisi constants to the largest satisfying fraction of MaxCut, Max $k$XOR, and Max $k$SAT \cite{dembo2017extremal, sen2017optimization, panchenko2017ksat}. We apply these techniques to explicitly compute the Parisi values for Max $k$XOR at small $k$, from which we find the optimal satisfying fraction of large random instances. Comparison to
our results for depth-$1$ QAOA and the threshold algorithm reveals still much room for improvement over these local algorithms (see Figure \ref{fig:parisicomparison} below).


\begin{figure}[htbp]
	\centering
	\includegraphics[width=5in]{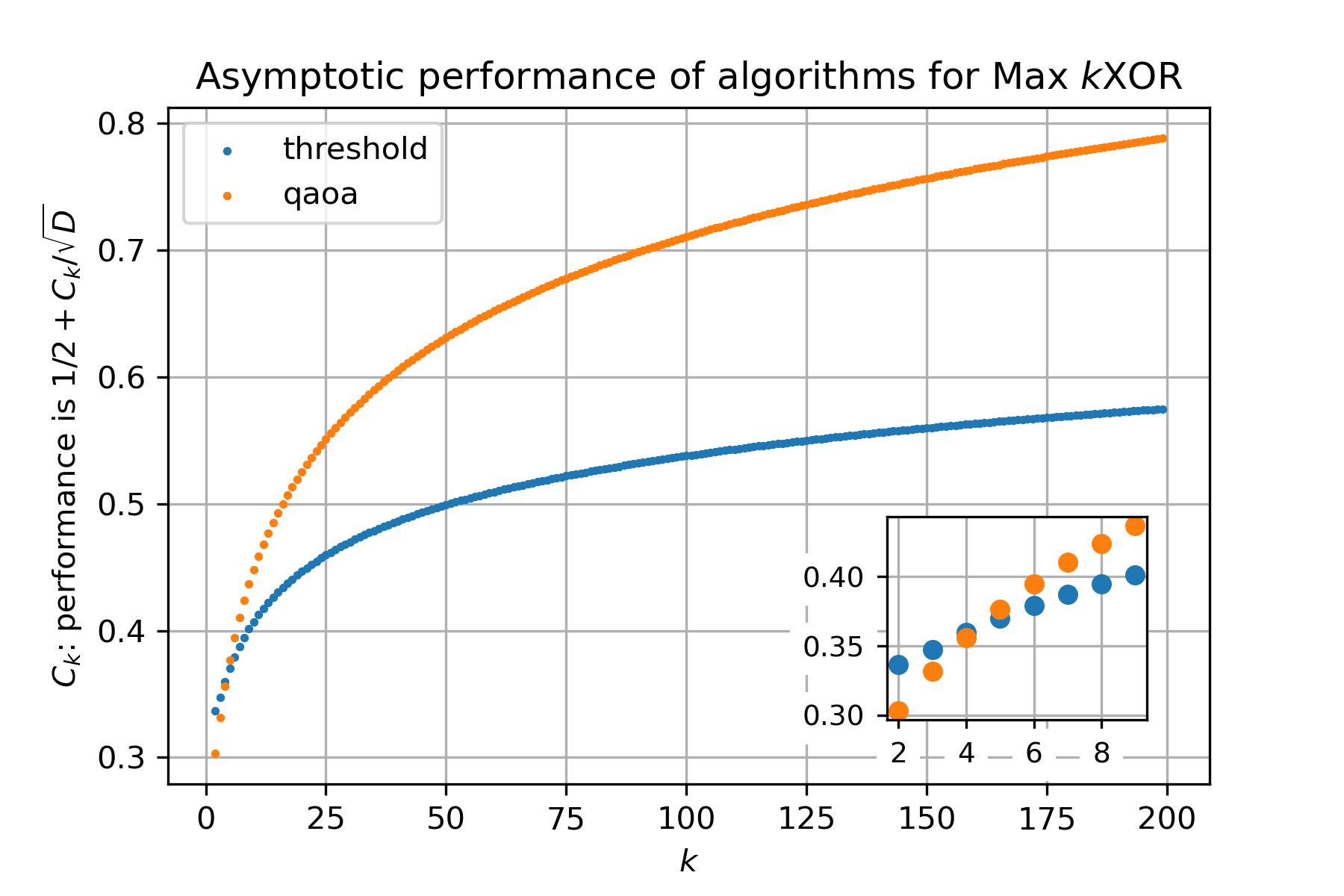}
	\caption{ Color online. Comparison of the satisfying fraction (performance) of the threshold algorithm and the depth-1 QAOA for Max $k$XOR at large $D$ (and $D \gg k$). The constant $C_k$ for QAOA grows more quickly with $k$ than for the threshold algorithm. Although the threshold algorithm outperforms the depth-1 QAOA for $k\in \{2,3,4\}$, at all other $k$, the QAOA is the better algorithm.}
	\label{fig:asymptoticcomparison}
\end{figure}

\begin{figure}[htbp]
	\centering
	\includegraphics[width=5in]{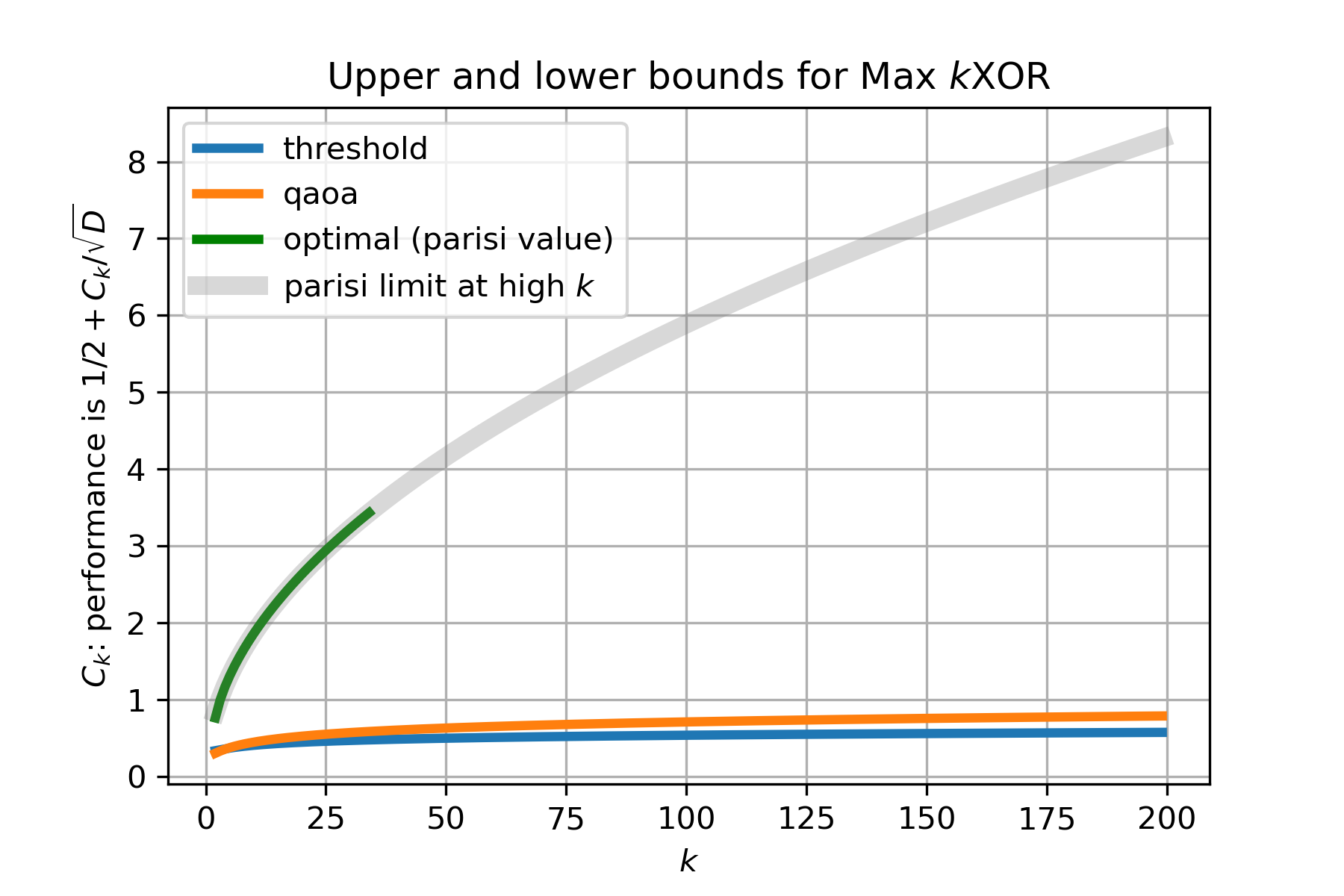}
	\caption{ Color online. A comparison of the satisfying fraction (performance) of the depth-1 QAOA and the threshold algorithm with the optimal value for large-degree, triangle-free instances of Max $k$XOR. We assume $D \gg k$. Since the Parisi value $P(k)$ converges to a constant as $k \to \infty$, the optimal satisfying fraction as $D\to\infty$ scales as $1/2 + \Theta(\sqrt{k/D})$ by \cite{sen2017optimization}. The one-local algorithms do not perform as well.}
	\label{fig:parisicomparison}
\end{figure}

\paragraph{Our contributions and guide to the paper.}
We first analyze both depth-1 QAOA, and a natural generalization
of the classical 
threshold algorithm \cite{hirvonen2017large}, on $(D+1)$-regular, triangle-free Max $k$XOR instances, in Sections \ref{sec:qaoa} and \ref{sec:threshold}, respectively. A comparison of the obtained performance results is shown in Figure~\ref{fig:asymptoticcomparison}, 
motivated by the algorithms' similar performance on MaxCut \cite{hastings2019classical, Marwaha2021localclassicalmax}.
In particular, we simplify 
the maximum satisfying fraction to a function of $k, D$, and a single algorithm parameter:
\begin{thm}
\label{thm:qaoa1}
The satisfying fraction of depth-1 QAOA on Max $k$XOR described by triangle-free, $D$-regular hypergraphs is
\begin{align*}
	  	 QAOA_1(\gamma,\beta) & = \frac{1}{2} -\frac{1}{4}is\big( (p+iqc^D)^k - (p-iqc^D)^k \big)
\end{align*}
where 
\begin{align*}
s = sin(\gamma), \; c = cos(\gamma), \; p = cos(2\beta), \; q = sin(2\beta).
\end{align*}
Furthermore, this holds if the hypergraphs are not triangle-free but the parities are chosen randomly.
\end{thm}

We optimize this expression at every $k < 200$, for each $D < 300$ and when $D\to\infty$.
The algorithm performance results also hold for arbitrary Max $k$XOR instances with random signs (parities).
Somewhat surprisingly, for $k > 4$ and at large enough~$D$, the depth-1 QAOA outperforms the threshold algorithm. This analysis verifies the asymptotic performance shown by \cite{hastings2019classical} for MaxCut (a special case of Max 2XOR), and extends and improves upon \cite{farhi2015quantum}'s $\beta=\pi/4$ analysis for Max 3XOR; 
see Appendix~\ref{sec:max3xorworked} for a detailed presentation of the $k=3$ case. 
See Section \ref{sec:comparison} for a comparison of the two algorithms at fixed $k$, including a table of optimal parameters at large degree.
\newline

We then identify the optimal satisfying fraction of large random instances for all $k$. We do this by conducting numerical calculations of the Parisi value $P(k)$ for all $k < 35$ in Section \ref{sec:parisinumerics}. See Figure \ref{fig:parisicomparison} for a comparison of these results with the algorithmic analysis.
Both algorithms are seen to perform far from optimally as $k$ increases. We observe that the value $P(k)$ converges to $\sqrt{2 \log{2}}$ (its large-$k$ limit) by $k=15$ (see Figure \ref{fig:parisivalue}).  We also show that $P(k)$ is at most $\sqrt{2 \log{2}}$ for all fixed $k$. 
To demonstrate the flexibility of the numerical calculation, we calculate the optimal satisfying fraction of large random $k$SAT instances, which is related to a similar spin glass model; see Appendix \ref{sec:ksat}. On Max 3SAT, the satisfying fraction given $m = \alpha n$ clauses is at most $7/8 + 0.278/\sqrt{\alpha}$ for high enough $\alpha$.
\newline

In Section \ref{sec:nlts}, we extend the NLTS result of \cite{bravyi2019obstacles} from quantum circuits with global bit-flip $X^{\otimes |V|}$ ($\mathbb{Z}_2$) symmetry, such as QAOA for MaxCut, to those where the circuit unitary commutes with an operator $X^{\otimes |V_+|}$ acting on a large, fixed subset $V_+ \subsetneq V$ of qubits. We call this \emph{partial $\mathbb{Z}_2$ symmetry}:
\begin{thm}[informal]
\label{ref:nltsthminformal}
There exists a family of Hamiltonians with partial $\mathbb{Z}_2$ symmetry that have no low-energy states accessible from low-depth quantum circuits with the same partial $\mathbb{Z}_2$ symmetry.
\end{thm}
The Hamiltonians we construct are derived from instances of the Max 3XOR problem. The QAOA unitary inherits the symmetry from the problem instance. This implies the following:
\begin{cor}[informal]
\label{ref:nltscorinformal}
There exists a family of fully-satisfiable Max 3XOR instances where the QAOA can only satisfy $99\%$ of clauses until depth at least logarithmic in the problem size.
\end{cor}
This is a constant-factor obstruction for low-depth QAOA. Note that we did not optimize the constant.
We conjecture this \emph{partial $\mathbb{Z}_2$ symmetry} can be similarly used to show an obstruction to QAOA for Max $k$XOR at each $k$. Our proof technique extends that of~\cite{bravyi2019obstacles} in a general way and may be useful for finding obstructions to QAOA on other classes of problems.  
\newline

We conclude the paper in Section~\ref{sec:conclusions} by summarizing our results and outlining some open research questions. 


\section{Depth-1 QAOA on Max $k$XOR}
\label{sec:qaoa}
The standard way to study Max $k$XOR on a quantum computer is to encode the problem as a $n$-qubit Hamiltonian. We use $X_i,Y_i,Z_i$ to denote the Pauli matrices acting on qubit $i$. The cost Hamiltonian for Max $k$XOR with $m$ clauses and $n$ variables can be written as $C = m/2 + \sum d_{n_1...n_k} Z_{n_1}...Z_{n_k} $, with each $d_{n_1...n_k}\in\{\pm1/2\}$ representing a clause.
Here, $C$ evaluates to the number of satisfied clauses on computational basis states $\ket{z_1 z_2 \dots z_n}$.
Since the QAOA's output is not deterministic, a common performance measure is the expected value, $\langle C\rangle$. (One can repeat the algorithm in order to achieve an outcome of at least $\langle C \rangle$ with high probability.)
So we consider the expected value of the depth-1 QAOA on each
$ d_{n_1...n_k}Z_{n_1}...Z_{n_k}$ term, as functions of the algorithm parameters. Previous analyses obtained exact formulas for $\langle C\rangle$ for the depth-1 QAOA for MaxCut \cite{Wang2018}, and 
bounds to the parameter-optimized $\langle C\rangle$ for Max 3XOR \cite{farhi2015quantum}.
\newline

We simplify the performance of the QAOA at depth-1 on triangle-free Max $k$XOR instances to a function of $k, D$, and a single hyperparameter. We optimize this expression at every $k < 200$, for each $D < 300$ and when $D\to\infty$.
\newline

In order to achieve this, we use the proof technique of \cite{Wang2018} (named the \emph{Pauli Solver} algorithm in \cite[Sec. 5.4.2]{hadfield2018quantum}). 
The main idea is that the expected value of each QAOA term can be written 
\begin{align*}
    \langle d_{n_1...n_k}Z_{n_1}...Z_{n_k} \rangle_1 
&= d_{n_1...n_k} \bra{\psi_0} \left(U_C^\dagger \left(U_B^\dagger Z_{n_1}...Z_{n_k} U_B\right) U_C\right) \ket{\psi_0} \\
&= d_{n_1...n_k} \Tr[\rho_0 \left(U_C^\dagger \left(U_B^\dagger Z_{n_1}...Z_{n_k} U_B \right)U_C\right)],
\end{align*}
which, as suggested by the parentheses, can be computed using a sequence of operator conjugations, followed by taking the initial state expectation value. 
Here $U_C=e^{-i\gamma C}$, $U_B=e^{-i\beta \sum_{j=1}^n X_j}$ and $\ket{\psi_0}=\ket{+}^{\otimes n}$.
For this standard initial state, operator terms containing any $Y$ or $Z$ operators give zero contribution to the initial state expectation value. The remaining terms can be collected after computing the final conjugated operator. 
This trick is particularly convenient for directly incorporating the effects of problem and algorithm locality.
\newline

We now outline the analysis of the depth-1 QAOA on regular, triangle-free instances of Max $k$XOR for general $k$.
We provide 
a detailed analysis for the $k=3$ case 
in Appendix~\ref{sec:max3xorworked}.
\newline

Consider a fixed $d_{n_1...n_k}Z_{n_1}...Z_{n_k}$ in $C$. Suppose each each node $i$ is involved in $D_i + 1$ clauses.
For the innermost conjugation by the mixing operator $U_B$, exponentials $e^{-i\beta X_j}$ with $j\notin \{n_1,\dots,n_k\}$ will commute through and cancel, to give $U_B^\dagger Z_{n_1}...Z_{n_k} U_B=e^{-2i\beta (X_{n_1}+\dots +X_{n_k})}Z_{n_1}...Z_{n_k}=\prod_{i=a}^k(pZ_i + qY_i)$ 
where $p = cos(2\beta)$ and $q = sin(2\beta)$. This can be seen from the properties of the Pauli matrices. 
This expression 
expands to  
a
binomial sum of terms with $j$-many $Y$s and $(k-j)$-many $Z$s, each weighted by $\binom{k}{j}q^jp^{k-j}$, for $j=0,1,\dots, k$.
\newline

Fix a term in the sum, and consider the subsequent conjugation by the phase operator $U_C$.
Exponentials in $U_C=\prod e^{-i\gamma d_{n_1\dots n_k}Z_{n_1}\dots Z_{n_k}}$ not involving qubits in $\{n_1,\dots,n_k\}$ again commute through and cancel. Let's look at the remaining exponentials. Because the instance is triangle-free, any exponential other than that of  $d_{n_1...n_k}Z_{n_1}\dots Z_{n_k}$ can only share one variable with the term (if it does, call it a \emph{neighboring} exponential), and it will not share variables with other neighboring exponentials. Because uncancelled $Z$'s give zero expectation, 
each of the $\sum_i D_i$ neighboring exponentials  will only contribute a factor of $c = cos(\gamma)$ (see  Appendix~\ref{sec:max3xorworked}). The central exponential will contribute $s = sin(\gamma)$ times a coefficient that depends on $j$. This is because $[Z,Y]=-2iX$, but [$Z\otimes Z,Y\otimes Y]=0$. If $j$ is even, then this contribution will be zero, and for odd $j=4\ell\pm 1$ the contribution will be $\pm 2d_{n_1...n_k}^2s = \pm 0.5s$.\footnote{The depth-1 QAOA has the same performance if the instance was not triangle-free but with random signs. Any neighboring exponentials that may interact will introduce uncancelled $d_{n_1'...n_k'}$, which gives zero expectation.}
\newline

Putting this together, the expectation is
\begin{align*}
	\langle d_{n_1...n_k}Z_{n_1}...Z_{n_k} \rangle_1 & = \frac{1}{2}s\sum_{j=1,3,5,...}^k q^j p^{k-j} (-1)^{(j-1)/2} \sum_{I \subseteq V; |I| = j} c^{\sum_{i\in I}D_i} .
\end{align*}
We emphasize that the right-hand size is independent of the set of $d_{n_1'...n_k'}$ values. This helps us bound the expected performance of each 
triangle-free instance.
\newline

The analysis simplifies when the instance is $(D+1)$-regular, i.e. each variable appears in exactly $D+1$ terms in $C$. In this case the expectation value of each clause is the same: $ \langle C \rangle/ m = \langle d_{n_1...n_k}Z_{n_1}...Z_{n_k} \rangle_1$. \newline

For each $j$, there are $jD$ neighbor clauses and ${k \choose j}$ terms of that type, so we have 
\begin{align*}
	\langle d_{n_1...n_k}Z_{n_1}...Z_{n_k} \rangle_1 & = \frac{1}{2}s\sum_{j=1,3,5,...}^k {k \choose j} q^j p^{k-j} (-1)^{(j-1)/2}c^{jD} 
	= -\frac{1}{2}si\sum_{j=1,3,5,...}^k {k \choose j} (iqc^D)^j p^{k-j}.
\end{align*}
Applying the binomial theorem,
this further simplifies to (the real function)
\begin{align*}
	  	\langle d_{n_1...n_k}Z_{n_1}...Z_{n_k}  \rangle_1 (\gamma,\beta) & = -\frac{1}{4}si\big( (p+iqc^D)^k - (p-iqc^D)^k \big) .
\end{align*}

This proves Theorem \ref{thm:qaoa1}. 
\newline

We optimize this expression at fixed $D$ and in the large-degree limit in the following subsection (\ref{sec:optimizeqaoakxor}). The code is \href{https://nbviewer.jupyter.org/github/marwahaha/QuAIL-2021/blob/main/maxkxor.ipynb}{online at this link}. In Figure \ref{fig:qaoamaxkxor} we show a plot of optimal values across $k$ and $D$. See Table \ref{tab:maxkxorlargedegree} for the optimal values at small values of $k$ in the large degree limit.
\newline

When $k=2$, the average value $\langle d_{ab} Z_a Z_b \rangle$ simplifies to $-0.25si(2ipqc^D) = 0.5spqc^D$. This exactly reproduces the depth-1 QAOA formula for MaxCut on regular, triangle-free instances, obtained in Corollary 1 of \cite{Wang2018}. On these instances, the depth-1 QAOA performs the same on Max 2XOR as it does for MaxCut.
\newline

We conjecture the optimal performance here is a lower bound on the performance for triangle-free graphs with maximum degree $D+1$.  See Appendix \ref{sec:max3xorworked} for a proof of this when $k=3$. For the general case this could plausibly be shown in two ways: either by analyzing QAOA directly for
instances with bounded degree, or by embedding the instance in a larger $(D+1)$-regular instance and using the analysis in this section.

\subsection{Optimizing the depth-1 QAOA on Max $k$XOR}
\label{sec:optimizeqaoakxor}

\begin{figure}[tbp]
	\centering
	\includegraphics[width=5in]{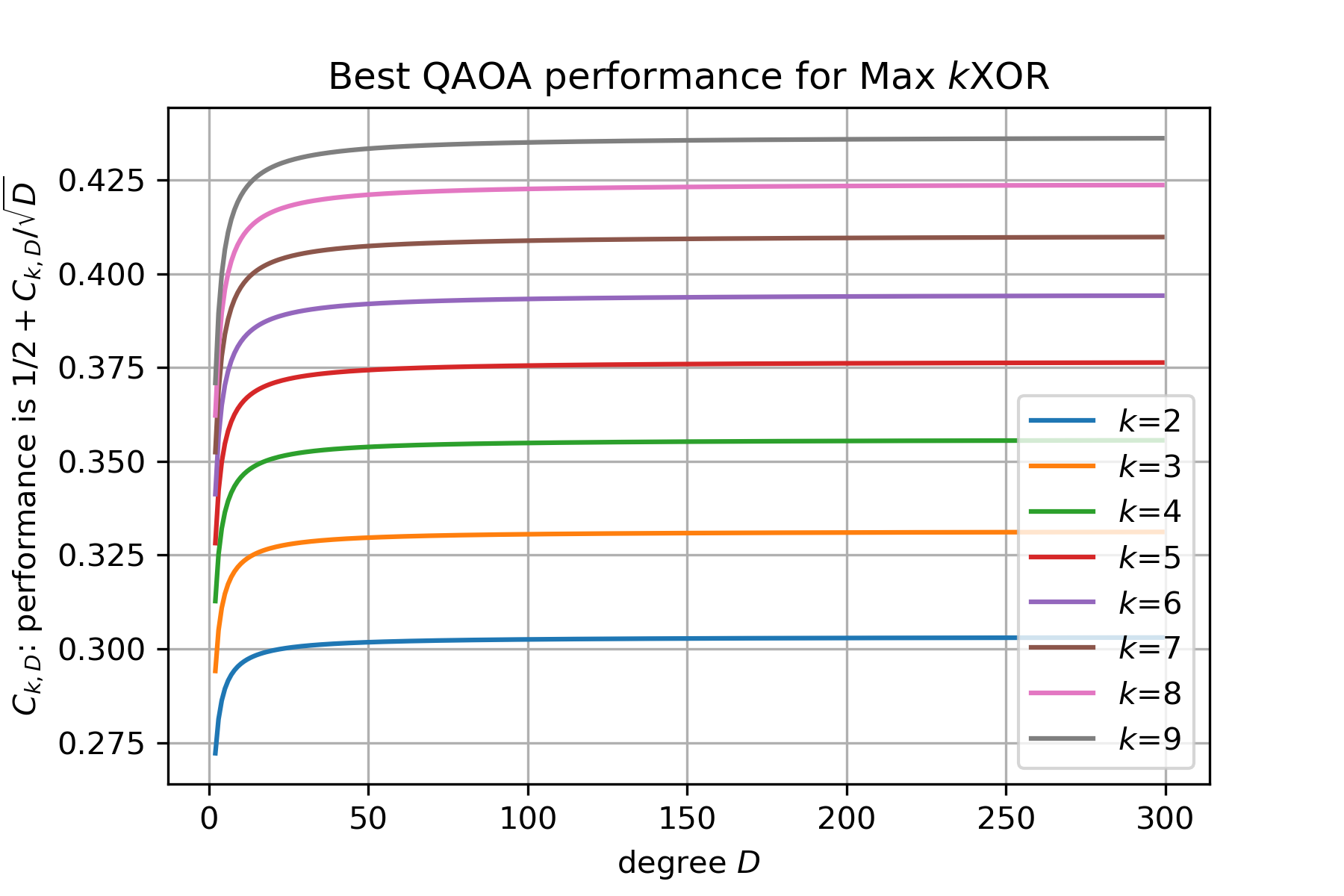}
	\caption{ Color online. The optimal performance of the depth-1 QAOA of any regular, triangle-free Max $k$XOR instance.
	Observe that the performance increases with $k$ for all $D$. }
	\label{fig:qaoamaxkxor}
\end{figure}

Consider the expression from before:
\begin{align*}
	  	\langle d_{n_1...n_k}Z_{n_1}...Z_{n_k} \rangle_1 & = -\frac{1}{4}si\big( (p+iqc^D)^k - (p-iqc^D)^k \big) 
\end{align*}
Don't let the imaginary terms confuse you, this is a real expression! We can take the derivative with respect to $\beta$ and find the maximum value:
\begin{align*}
	\dbyd{\beta} = 0                    & \xrightarrow[]{}
	& \dbyd{\beta} (p+iqc^D)^k            & = \dbyd{\beta} (p-iqc^D)^k
	\\
	  &   & k(p+iqc^D)^{k-1}(p'+iq'c^D)     & = k(p-iqc^D)^{k-1}(p'-iq'c^D)                                                                                                        
	\\
	  &   & (\frac{p+iqc^D}{p-iqc^D})^{k-1} & = \frac{p'-iq'c^D}{p'+iq'c^D} = \frac{-q-ipc^D}{-q+ipc^D} = \frac{-iqc^{-D} + p}{-iqc^{-D} - p} = -\frac{p - iqc^{-D}}{p + iqc^{-D}} 
\end{align*}
And the maximum value when taking the derivative with respect to $\gamma$:
\begin{align*}
\dbyd{\gamma} &= 0                         \xrightarrow[]{}
	&   0 = -0.&25ci\big( (p+iqc^D)^k - (p-iqc^D)^k \big)
	\\
	  &   &   & -0.25si\big(                                       k(p+iqc^D)^{k-1}(iqDc^{D-1}(-s)) - k(p-iqc^D)^{k-1}(-iqDc^{D-1}(-s)) \big) 
\end{align*}
This implies the following:
\begin{align*}
	\big( (p+iqc^D)^k - (p-iqc^D)^k \big)                 & = s^2kiqDc^{D-2}\big( (p+iqc^D)^{k-1} + (p-iqc^D)^{k-1} \big)     
	\\
	(p+iqc^D)^{k-1}\big( p + iqc^D - is^2kqDc^{D-2} \big) & = (p-iqc^D)^{k-1}\big( p - iqc^D + is^2kqDc^{D-2} \big)           
	\\
	(\frac{p+iqc^D}{p-iqc^D})^{k-1}                       & = \frac{ p - iqc^D + is^2kqDc^{D-2}}{ p + iqc^D - is^2kqDc^{D-2}} 
	= \frac{ p + iqc^{D-2}(s^2kD - c^2)}{p - iqc^{D-2}(s^2kD - c^2)}
\end{align*}
Setting the two equations equal to each other, where $\zeta = s^2 kD - c^2$:
\begin{align*}
	  &          & \frac{ p + iqc^{D-2}\zeta}{p - iqc^{D-2}\zeta}     & = -\frac{p - iqc^{-D}}{p + iqc^{-D}}                               
	\\
	  & \implies & p^2 + ipq(c^{-D} + c^{D-2}\zeta) - q^2 c^{-2}\zeta & = -\Big( p^2 - ipq(c^{-D} + c^{D-2}\zeta) - q^2c^{-2}\zeta \Big) 
	\\
	&\implies & 
	2(p^2 - q^2c^{-2}\zeta) & = 0
	\\
	  & \implies & p^2/q^2                                              & = 1/q^2 - 1 = \zeta/c^2                                           \\
	  & \implies & q                                                    & = c/s \sqrt{1/kD}                                                  
\end{align*}
This tells us how $\gamma$ and $\beta$ are related in the optimal solution. So now we have a single-variable function that we can optimize. This gives an expression for every $k$ and $D$. See Figure \ref{fig:qaoamaxkxor}.
\newline

As a sanity check, we show this constraint is satisfied for known results. At $k=2$, \cite{Wang2018} shows that when $\beta = \pi/8$, $q = sin(2\beta) = 1/\sqrt{2}$, which gives $s/c = tan(\gamma) = 1/\sqrt{D}$. At $k=3$ at large degree, $\gamma = t/\sqrt{D}$ gives $sin(2\beta) = 1/(t\sqrt{k})$, and $t \approx 1.0535$ and $\beta \approx 0.29$ satisfies this equation, as described in Appendix \ref{sec:max3xorworked}.
\newline

At $D$ large, we can take a small-angle approximation of $\gamma$. For example, if $\gamma = t/\sqrt{D}$ then $q = sin(2\beta) = 1/(t\sqrt{k})$, then we can compute the limiting behavior of QAOA. Then $c = cos(\gamma) \approx 1 - \gamma^2/2$, so $c^D = e^{-t^2/2}$.
\begin{align*}
	\langle d_{n_1...n_k} Z_{n_1}...Z_{n_k} \rangle_1 & \approx -0.25 \gamma i ((p + iqe^{-t^2/2})^k - (p-iqe^{-t^2/2})^k) 
	\\
	                                   & = \frac{1}{\sqrt{D}}(-0.25ti)                                      
	\Big( ((1-1/(t^2k))^{0.5} + i\frac{1}{t\sqrt{k}}e^{-t^2/2})^k - ((1-1/(t^2k))^{0.5} - i\frac{1}{t\sqrt{k}}e^{-t^2/2})^k \Big)
	\\
	                                   & = \frac{-0.25i}{\sqrt{D}}t^{1-k} k^{-k/2} \Big(                    
	(\sqrt{t^2k-1} + ie^{-t^2/2})^k
	- (\sqrt{t^2k-1} - ie^{-t^2/2})^k
	\Big)
\end{align*}
Optimizing this value  over $t$ gives the limiting behavior at large $D$. See Table \ref{tab:maxkxorlargedegree} for a numerical calculation of $C$ in the large degree limit.


\section{A local threshold algorithm on Max $k$XOR}
\label{sec:threshold}

Recall the threshold algorithm of \cite{hirvonen2017large} for MaxCut.
This is also a one-local algorithm; given an initial assignment, the output assignment for each vertex depends only on the initial values assigned to its neighbors. 
\cite{hastings2019classical} showed that 
a generalization of this algorithm outperformed the depth-1 QAOA on MaxCut for all regular triangle-free graphs.
\newline

The threshold algorithm naturally generalizes to 
Max $k$XOR as follows. Each node makes an update decision based on its 
and its neighbors' current assignments.\footnote{Each node also needs the values $d_{n_1...n_k} = \pm 1/2$ associated with the clauses it is involved in. This is distinct from MaxCut, where the $d_{ab}$ are all the same.} Given a parameter $\mu$:
\begin{enumerate}
	\item Assign every node a value $\pm 1$ with equal probability, independently.
	\item Simultaneously, every node checks the number of clauses it is involved in that are satisfied; if it is at most $\mu$ then the node's value is switched.\footnote{\cite{hirvonen2017large} considers a threshold $\tau = D+1-\mu$ with respect to our notation.}
\end{enumerate}

We extend the analysis of \cite{hirvonen2017large} to find an expression for the performance of this algorithm, given any $(D+1)$-regular, triangle-free instance. We numerically optimize this expression at every $k < 200$, for each $D < 300$ and in the large-degree limit.
\newline

Consider a fixed clause. As it's triangle-free, each of the neighbor clauses are i.i.d.\ satisfied with probability $1/2$ after Step 1. This is because in each adjacent clause there are $k-1$ i.i.d remaining variables that could be $\pm 1$.\footnote{The adjacent clauses are also i.i.d.\ when the clauses are arbitrary but the clause signs are random. As a result, this analysis is exactly the same if the instance is \emph{not} triangle-free but the clause signs are randomly chosen.}  Let's say the fixed clause is unsatisfied. Then the chance one of its nodes (a \emph{member node}) will flip is the chance that at most $\mu$ of the clauses containing that node are satisfied. If instead the fixed clause is satisfied, then a member node will flip if at most $\mu -1$ neighbor clauses are satisfied (because one is already satisfied). This difference is the secret to the algorithm: the flip operation has a slight bias towards satisfying more clauses. Hence we have the quantities: 
\begin{align*}
	g &= \Pr[flip | \text{UNSAT}] = \frac{1}{2^D}\sum_{i=0}^\mu {D \choose i}  & h &=1-g \\
	r &= \Pr[flip | \text{SAT}] = \frac{1}{2^D}\sum_{i=0}^{\mu-1} {D \choose i} & s &=1-r   & 	g - r = \Delta &=  \frac{1}{2^D} {D \choose \mu}
\end{align*}
If a clause is initially unsatisfied, it needs an odd number of variables to flip to become satisfied (these correspond to odd terms of a binomial with probability $g$). If it is initially satisfied, it needs an even number of flips out of $k$ to remain so (these correspond to even terms of binomial with probability $r = g - \Delta$). Altogether, the chance $F = 1/2 + C_{k,D}/\sqrt{D}$ of satisfying the clause is:
\begin{align*}
	F & = \frac{1}{2} \Big(\sum_{j=1,3,5,...}^k {k \choose j} g^j h^{k-j} + \sum_{j=0,2,4,...}^k {k \choose j} r^j s^{k-j} \Big) 
	\\
	  & = \frac{1}{4}(1 - (h-g)^k) + \frac{1}{4}(1 + (s-r)^k)      
	   = \frac{1}{2} + \frac{(1-2g+2\Delta)^k - (1-2g)^k}{4}        
\end{align*}
Then, $\mu$ can be optimized at each $k$ and $D$ to find the best threshold value. When $k=2$, the expression simplifies to $1/2 + \Delta(1-2g + \Delta)$. This matches the expression given in Lemma 3 of \cite{hirvonen2017large} for $k=2$. 
When $k$ is odd, $F$ is unchanged when switching $h$ and $r$; so there are two optimal thresholds, symmetric around $D/2$.
\newline

At large degree, the asymptotic value of the constant $C_k = \lim_{D\to\infty} C_{k,D}$ and optimal $\mu = D/2 + \alpha\sqrt{D}$ can also be numerically computed:
\begin{align*}
    \Delta &\approx \sqrt{\frac{2}{\pi D}} e^{-2\alpha^2} &    &g\approx \Phi(2\alpha) + \frac{\Delta}{2} = \frac{1}{2}(\erf(\alpha\sqrt{2}) + 1 + \Delta)
\end{align*}
\begin{align*}
	F - \frac{1}{2} &\approx \frac{(\erf(-\alpha \sqrt{2}) + \Delta)^k - (\erf(-\alpha \sqrt{2}) - \Delta)^k}{4}  
\approx \frac{k\Delta}{2}\erf^{k-1}(-\alpha \sqrt{2}).                                            
\end{align*}
The performance is maximized when $(k-1)\Delta \sqrt{D} = 2\alpha \erf(\alpha\sqrt{2})$. A numerical computation can be done for each $k$; values for low $k$ are given in Table \ref{tab:maxkxorlargedegree}. At larger $k$, the optimal $|\alpha|$ increases, which increases the optimal value $C_k$. As noted in \cite{hastings2019classical, Marwaha2021localclassicalmax}, there are oscillations at finite degree because the optimization is over a discrete threshold parameter. The $C_{k,D}$ are not necessarily strictly increasing at fixed $D$, even if the $C_k$'s are. See Figure \ref{fig:comparison} and Table \ref{tab:maxkxorlargedegree} for a detailed comparison with depth-1 QAOA.
\newline

As noted in \cite{hirvonen2017large}, this algorithm can be used for graphs that have at most $D+1$ clauses per variable. One can ``simulate'' the other neighbors and use the same algorithm. So this analysis is a lower bound for the satisfying fraction on bounded-degree, triangle-free instances.



\section{Comparison of one-local algorithms}
\label{sec:comparison}

\begin{figure}[htbp]
	\centering
	\includegraphics[height=8in]{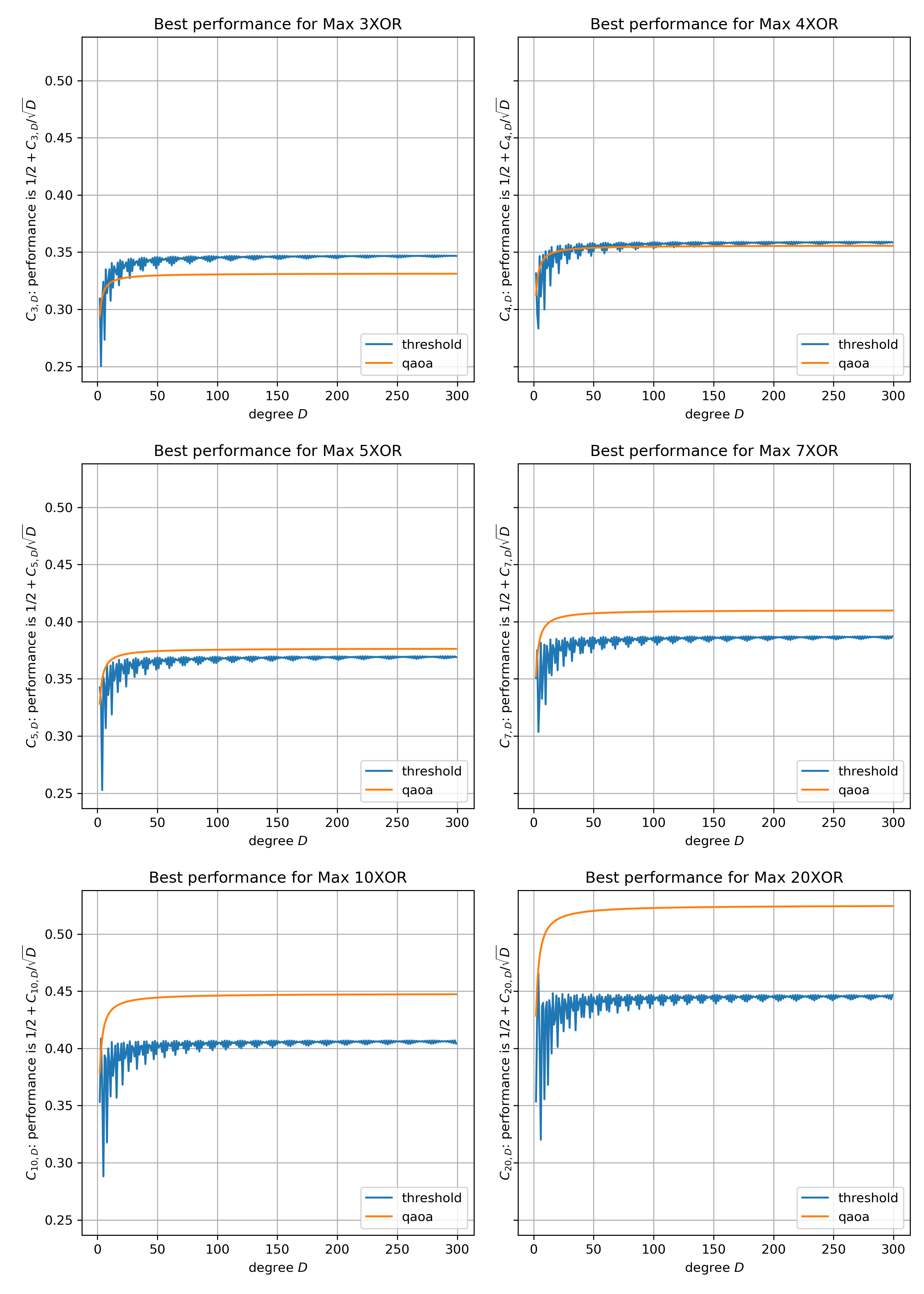}
	\caption{ Color online. Comparison of the threshold algorithm and the depth-1 QAOA on Max $k$XOR. Both algorithms have the optimal scaling of $1/2 + \Theta(1/\sqrt{D})$. The threshold algorithm has oscillations because it is optimized over a discrete parameter choice. For $k > 4$, QAOA outperforms the threshold algorithm for most finite values of $D$ and in the large-degree limit.}
	\label{fig:comparison}
\end{figure}

See Figure \ref{fig:comparison} for a comparison of the two algorithms across $k$ and $D$, and Figure \ref{fig:asymptoticcomparison} for a comparison at the large-degree limit (i.e. of $C_k$). See also Table \ref{tab:maxkxorlargedegree} for a listing of optimal large-degree values at small $k$. The code is \href{https://nbviewer.jupyter.org/github/marwahaha/QuAIL-2021/blob/main/maxkxor.ipynb}{online at this link}.
\newline

\begin{table}[htb]
	\centering
	\begin{tabular}{|c||c|c|c||c|c|}
		\hline
		$k$ & $C_{QAOA_1}$ & $\gamma \sqrt{D}$ & $\beta$ & $C_{\mu}$ & $\alpha$ \\ \hline
		2   & 0.30326      & 1.00001           & 0.39269 & 0.33649    & -0.43845 \\
		3   & 0.33146      & 1.05351           & 0.29000 & 0.34754    & -0.56611 \\
		4   & 0.35594      & 1.09779           & 0.23644 & 0.35948    & -0.64611 \\
		5   & 0.37671      & 1.13477           & 0.20254 & 0.37008    & -0.70408 \\
		6   & 0.39459      & 1.16637           & 0.17879 & 0.37934    & -0.74931 \\
		7   & 0.41025      & 1.19393           & 0.16105 & 0.38748    & -0.78625 \\
		8   & 0.42415      & 1.21833           & 0.14721 & 0.39471    & -0.81739 \\
		9   & 0.43665      & 1.24021           & 0.13605 & 0.40120    & -0.84426 \\
		10  & 0.44799      & 1.26005           & 0.12683 & 0.40707    & -0.86783 \\
		11  & 0.45837      & 1.27817           & 0.11906 & 0.41243    & -0.88881 \\
		12  & 0.46793      & 1.29485           & 0.11241 & 0.41736    & -0.90769 \\
		13  & 0.47679      & 1.31031           & 0.10664 & 0.42192    & -0.92483 \\
		14  & 0.48505      & 1.32469           & 0.10157 & 0.42615    & -0.94052 \\
		15  & 0.49279      & 1.33815           & 0.09708 & 0.43010    & -0.95497 \\
		16  & 0.50005      & 1.35081           & 0.09307 & 0.43381    & -0.96836 \\
		17  & 0.50690      & 1.36273           & 0.08946 & 0.43729    & -0.98083 \\
		18  & 0.51338      & 1.37399           & 0.08619 & 0.44058    & -0.99249 \\
		19  & 0.51953      & 1.38469           & 0.08322 & 0.44370    & -1.00344 \\
		\hline
	\end{tabular}
	\caption{ The maximum satisfying fraction from the expected performance of the depth-1 QAOA ($C_{QAOA_1}$) and the threshold algorithm ($C_{\mu}$) on Max $k$XOR in the large-degree limit. The fraction has value $1/2 + C/\sqrt{D}$. The optimal $\gamma\sqrt{D}$ and threshold parameter $\alpha = (\mu-D/2)/\sqrt{D}$ are also listed. All values are truncated.
	The values at $k=2$ reproduce previously known results in \cite{Wang2018, hastings2019classical}. Note that for $k>4$, the depth-1 QAOA outperforms the threshold algorithm. The optimal $|\alpha|$ increases with $k$. The code is \href{https://nbviewer.jupyter.org/github/marwahaha/QuAIL-2021/blob/main/maxkxor.ipynb}{online at this link}.
	}
	\label{tab:maxkxorlargedegree}
\end{table}

On Max 3XOR, the depth-1 QAOA beats the threshold algorithm for $D \in \{3,  4,  6,  8, 11, 13, 18, 20, 27\}$. But as $D$ increases, the threshold algorithm performs better. The threshold algorithm has the same ``oscillations'' noted in \cite{hastings2019classical} and \cite{Marwaha2021localclassicalmax}, likely because of the optimization over discrete threshold $\mu$. We expect that slight adjustments to the threshold algorithm for these choices of degree may outperform the depth-1 QAOA, as in \cite{hastings2019classical}.
\newline

However, for larger $k$, the QAOA starts to outperform the threshold algorithm.  As noted in Table~\ref{tab:maxkxorlargedegree}, the depth-1 QAOA has higher performance starting at $k=5$. On these instances, the depth-1 QAOA is on average a better algorithm than the local classical algorithm described above. The threshold algorithm is an instance of a local tensor algorithm as defined by \cite{hastings2019classical}; it is possible that a different local tensor algorithm performs as well as the QAOA at higher $k$.

\section{An upper bound via numerical evaluation of the Parisi formula}
\label{sec:parisinumerics}
The optimal satisfying fraction of a sparse
random regular MAX $k$XOR problem is related to the ground state energy density $P(k)$ of a pure $k$-spin glass, as shown by \cite{sen2017optimization} (extending \cite{dembo2017extremal}).
At large degree, the satisfying fraction is $\frac{1}{2} + \frac{P(k)}{2} \sqrt{\frac{k}{D}}$ for nearly all instances.
At $k=2$, this recovers the famous Parisi constant $P_* = P(2)/ \sqrt{2} \approx 0.7632$ \cite{Crisanti_2002, panchenko2014introduction, alaoui2020algorithmic}, found in the expression for the optimal cut fraction of a sparse random graph at large degree.
\newline

A fully-connected, pure $k$-spin glass (also called \emph{the mean-field $k$-spin model}) has a Hamiltonian
\begin{align*}
    H_k(\vec{x}) = \lim_{N\to\infty} H_{k,N}(\vec{x}) = \frac{1}{N^{(k-1)/2}} \sum_{n_1,...,n_k = 1}^N g_{n_1,...,n_k} x_{n_1}...x_{n_k} ,
\end{align*}
where each $g$ is an i.i.d.\ standard Gaussian random variable. The ground state energy density $P(k)$ is then $\lim_{N\to\infty} \frac{1}{N} \max_{\vec{x}} H_{k,N}(\vec{x})$.
This value can be calculated via the Parisi formula.
Using a recent variational method from \cite{auffinger2016parisi}, this formula can be evaluated directly at zero temperature. However, the numerical values have only been calculated for $k=2$ (which is proportional to the Sherrington-Kirkpatrick model) and $k=3$ \cite{alaoui2020algorithmic}.
\newline

\begin{figure}[htbp]
	\centering
	\includegraphics[width=5in]{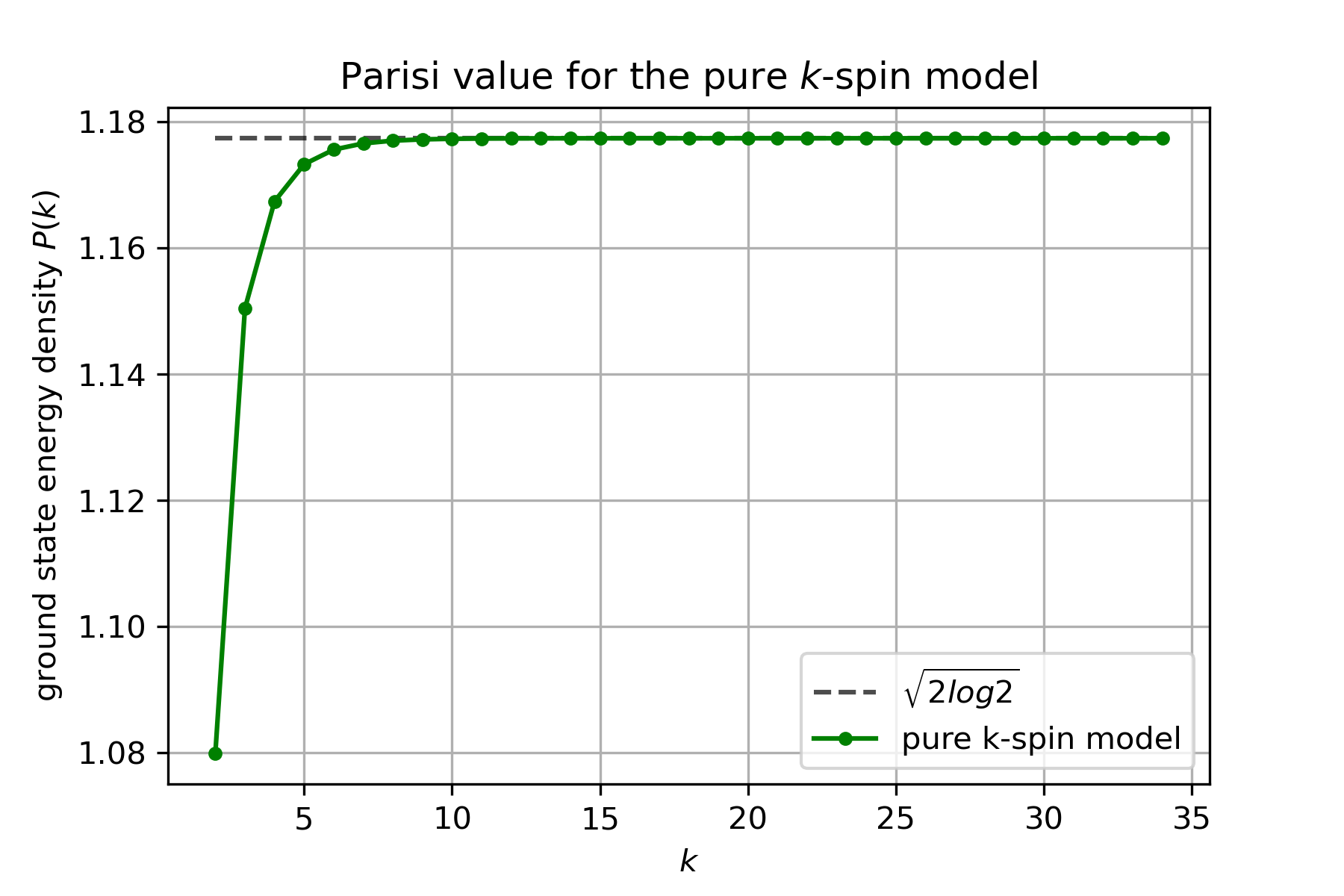}
	\caption{ The ground state energy density $P(k)$ of the pure $k$-spin model. As $k$ increases, $P(k)$ approaches the ground state energy density of the random energy model, with extremely close agreement by $k=15$.}
	\label{fig:parisivalue}
\end{figure}

We numerically minimize the Parisi formula for the pure $k$-spin glass for all $2 \le k < 35$. This reproduces the known numerical minima at $k=2$ and $k=3$. The code is \href{https://nbviewer.jupyter.org/github/marwahaha/QuAIL-2021/blob/main/parisi.ipynb}{online at this link}; it can be used to find the minimum energy density of any mean-field (fully-connected) spin glass.
\newline

\begin{table}[htbp]
\centering
\begin{tabular}{|c|c|c|}
\hline
$k$           & Known value     & calculated $P(k)$                \\ \hline
2   & 1.07928 & 1.0799   \\ \hline
3  & 1.150 & 1.1504    \\ \hline
4   & - & 1.1674  \\ \hline
5   & - & 1.1732   \\ \hline
6   & - & 1.1756   \\ \hline
7   & - & 1.1765  \\ \hline
8   & - & 1.1770  \\ \hline
9   & - & 1.1772  \\ \hline
10  & - & 1.1773 \\ \hline
11  & - & 1.1773  \\ \hline
12  & - & 1.1773  \\ \hline
13  & - & 1.1774  \\ \hline
14  & - & 1.1774  \\ \hline
15  & - & 1.1774  \\ \hline
\end{tabular}
\caption{ Listing of calculated Parisi values $P(k)$ for the pure $k$-spin glass describing Max $k$XOR at small $k$. The optimal satisfying fraction of a sparse random instance is $\frac{1}{2} + \frac{P(k)}{2}\sqrt{\frac{k}{D}}$ as the number of clauses per variable $D$ grows large. All values are truncated. If desired, our code can be run with better accuracy using more memory. The known values come from \cite{alaoui2020algorithmic} (they calculate $P(k)/\sqrt{2}$). Notice how the values converge to $\sqrt{2 \log{2}} \approx 1.17741$ by $k=15$. }
\label{tab:parisi}
\end{table}

The Parisi minimization procedure chooses the best piecewise-constant function as input. The approximation to the true value $P(k)$ improves as the number of pieces increases. Unlike \cite{alaoui2020algorithmic}, we do not explicitly calculate the derivative; however, we expect convergence because the Parisi functional has a unique minimizer \cite{Auffinger_2014}. Using 2 nonzero pieces and a spatial grid of 400 points per dimension\footnote{With $\mu$ nonzero pieces, the calculation uses $\mu + 1$ dimensions.}, the code ran on a mid-range laptop in about twenty minutes per value of $k$.
\newline

See Table \ref{tab:parisi} for a listing of values at small $k$. Our values agree with \cite{alaoui2020algorithmic} for $k=2$ and $k=3$ to within $0.1\%$ relative accuracy.
At high $k$, the computed constant $P(k)$ recovers its high-$k$ limit $\sqrt{2 \log{2}}$, the ground state energy density of the random energy model \cite{Derrida_PhysRevB.24.2613}. See Figure \ref{fig:parisivalue} for a plot across $k$. $P(k)$ is upper-bounded by its limit as $k\to\infty$; we give an informal proof of this in the following subsection (\ref{sec:pkupperbound}).
\newline

We compare this calculation of the optimal satisfying fraction to the performance of the algorithms we analyze.
This is possible because the induced subgraph around any edge of a random regular instance of Max $k$XOR is asymptotically always triangle-free with respect to the number of variables.
As a result, the expected satisfying fraction of a local algorithm is the same on triangle-free instances and sparse random regular instances with the same degree.
The separation between optimal value and algorithm performance increases with $k$; see Figure \ref{fig:parisicomparison}, which indicates much room for potential algorithm improvement. We emphasize that deeper quantum circuits (i.e., QAOA with depth $p>1$) may perform closer to optimally, but rigorous performance bounds in this setting appear difficult to calculate.
\newline

To demonstrate the
utility of this Parisi calculator, we also apply it to the Max $k$SAT problem; see Appendix \ref{sec:ksat} for details.
	
	\subsection{$P(k)$ is upper-bounded by its limiting value}
\label{sec:pkupperbound}

We use the same definitions to solve the Parisi functional as in many papers \cite{auffinger2013properties, panchenko2014introduction, Auffinger_2014, auffinger2016parisi, panchenko2017ksat}. Consider a step function on $[0,1)$ with two pieces: $0$ on $[0, \epsilon)$ and $m_1$ on $[\epsilon, 1)$ where $\epsilon \ll 1/k \le 1$. For a pure $k$-spin glass, $\xi(s) = s^k$. From Equation 4 of \cite{auffinger2016parisi}, the Parisi functional is then:
	$$
	P(k) = \Psi_0(0) - B = \Psi_0(0) - \frac{1}{2} \int_0^1 f(t) \xi^{''}(t) t dt
	$$
	The penalty term $B$ can be simplified:
	\begin{align*}
		B & = \frac{1}{2} m_1 \int_{\epsilon}^1 k(k-1)t^{k-1} dt             
		\\
		  & =\frac{(k-1) m_1}{2} (1 - \epsilon^k) \approx \frac{m_1(k-1)}{2} 
	\end{align*}
	Let's define $a_i = \sqrt{\xi^{'}(q_{i+1}) - \xi^{'}(q_i)}$, where $q_0 = 0, q_1 = \epsilon, q_2 = 1$. So $a_1 = \sqrt{k} \sqrt{1 - \epsilon^{k-1}} \approx \sqrt{k}$, and $a_0 = \sqrt{k} \sqrt{\epsilon^{k-1}} \approx 0$. We use these variables to recursively calculate $\Psi_i$, as in Equation 12 of \cite{auffinger2013properties}.
	\newline

	Define $\Psi_2(x) = |x|$ as the boundary condition as in \cite{auffinger2016parisi}, and $\Psi_0(x) = \E[\Psi_1(x + a_0 z)]$, averaging across a standard normal variable $z$.
	But since $a_0 \approx 0$ we have $\Psi_0(x) \approx \Psi_1(x)$.
	\newline

	We define $\Psi_1(x)$ as the following:
	\begin{align*}
		\Psi_1(x) & = \frac{1}{m_1} \log \Bigg(       
		\E \Big[ e^{m_1 \Psi_2(x + a_1 z)} \Big]
		\Bigg)
		\\
		          & \approx \frac{1}{m_1} \log \Bigg( 
		\E \Big[ e^{m_1 |x + \sqrt{k} z|} \Big]
		\Bigg)
	\end{align*}
	Then we can solve for $\Psi_0(0)$:
	\begin{align*}
		\Psi_0(0) \approx \Psi_1(0) & \approx \frac{1}{m_1} \log \Bigg(            
		\E \Big[ e^{m_1 \sqrt{k} |z|} \Big]
		\Bigg)
		= \frac{1}{m_1} \log \Bigg(
		e^{m_1^2 k / 2} \Big( \erf(m_1 \sqrt{k/2}) + 1 \Big)
		\Bigg)
		\\
		                            & = \frac{m_1 k}{2} + \frac{1}{m_1} \log \Big( 
		\erf(m_1 \sqrt{k/2}) + 1
		\Big)
		\le \frac{m_1 k}{2} + \frac{\log{2}}{m_1}
	\end{align*}
	This upper bounds the Parisi functional:
	\begin{align*}
		P(k) & = \Psi_0(0) - B           \lesssim \frac{m_1}{2} + \frac{\log{2}}{m_1} 
	\end{align*}
	The upper bound is minimized at $m_1 = \sqrt{2 \log{2}}$ with $P = \sqrt{2 \log{2}}$. So this means that $P(k)$ is upper bounded at approximately $\sqrt{2 \log{2}}$. We also observe this in our numerical calculations; see Figure \ref{fig:parisivalue}. At high $k$, the pure $k$-spin glass becomes the random energy model where $J = \sqrt{2}$, which has exactly this ground state energy density.

	
	\section{An NLTS-based obstruction for QAOA on Max $3$XOR}
	\label{sec:nlts}
	Here we construct Hamiltonians for Max $3$XOR that satisfy the No Low-energy Trivial States (NLTS) property for quantum circuits with a particular symmetry property. Our motivation is that QAOA circuits at odd $k$ do not have the global bit-flip $\mathbb{Z}_2$ symmetry considered in \cite{bravyi2019obstacles}; rather, we explain how QAOA circuits in this case satisfy a property we call partial $\mathbb{Z}_2$ symmetry. We use this to show a constant-fraction obstruction for low-depth QAOA on Max 3XOR, analogous to the previous results obtained for MaxCut. 
	\newline

	Given a unitary operator $A$, a family of quantum circuits $\{U_\alpha\}$ is said to have $A$ as a symmetry if each circuit satisfies $A^\dagger U_\alpha A=U_\alpha$~(see e.g.~\cite{shaydulin2020classical}).
	For Max $k$XOR with even $k$, QAOA circuits have a so-called\footnote{As $A^2=I$, 
	\{I,A\} gives a representation of the two-element group $\mathbb{Z}_2$.} $\mathbb{Z}_2$ symmetry for $A=X_1X_2\dots X_n$, which corresponds to the classical operation of flipping all bits. Since the standard initial state  satisfies $A\ket{+}^{\otimes n}=\ket{+}^{\otimes n}$, the probability of measuring any particular bit string after running one of these QAOA circuits is the same as that of its complement.
	\newline

	The NLTS property holds for a family of Hamiltonians when the 
	quantum circuits and product states have $\mathbb{Z}_2$ symmetry~\cite{bravyi2019obstacles}. These Hamiltonians represent MaxCut instances on expander graphs. We generalize this result to construct a family of Max 3XOR instances, which have a slightly more general symmetry that also applies to the corresponding QAOA circuits. Let's call this \emph{partial $\mathbb{Z}_2$ symmetry}. This refers to when the unitary has $\mathbb{Z}_2$ bit-flip symmetry 
	with respect to some large, fixed subset $V_+ \subsetneq V$, where both $V/V_+$ and $V_+$ have cardinality $\Theta(|V|)$, and $|V/V_+| \ll |V_+|$. To be explicit, let's suppose $|V/V_+| < |V_+|/r$ for some constant $r > 4$. Importantly, the QAOA circuit for Max 3XOR has the same \emph{partial $\mathbb{Z}_2$ symmetry} on these instances, and so is obstructed from optimally solving these instances at sub-logarithmic depth.
	
	\begin{thm}[Formal version of Theorem \ref{ref:nltsthminformal}] \label{ref:nltsthmformal}
	There exists an $\epsilon > 0$ and a family of Hamiltonians $\{H_n\}$ with partial $\mathbb{Z}_2$ symmetry, each with ground state energy $0$, such that when $n > 5184 \cdot 8^d$, every depth-$d$ unitary $U$, composed of 2-qubit gates and with the same partial $\mathbb{Z}_2$ symmetry, gives $\langle +^{\otimes n} | U^\dagger H_n U | +^{\otimes n} \rangle > \epsilon n$.
	\end{thm}
	
	At a high level, we choose a large random MaxCut instance (with $\mathbb{Z}_2$ symmetry). We then add a small number of nodes, and convert every edge into hyperedges including one of the added nodes. The Hamiltonian then has a large ``core'' of nodes that obey $\mathbb{Z}_2$ symmetry. The QAOA circuits that encode these Hamiltonians also have this generalized symmetry:
	\begin{cor}[Formal version of Corollary \ref{ref:nltscorinformal}] \label{ref:nltscorformal}
	There exists a family of Hamiltonians $\{H_n\}$ representing Max $3$XOR instances that are fully satisfiable, but the QAOA can only achieve a satisfying fraction at most 
	\begin{align*}
	    0.99 + \frac{\sqrt{D-1}}{50D}
	    + o\Big(\frac{1}{\sqrt{D}}\Big)
	\end{align*}
	when the depth $p < \log_2 (n/5184)(648D)^{-1}$.
	\end{cor}
	
	This is a constant-fraction obstruction to QAOA on Max 3XOR. It may be possible to improve the constant; we did not optimize it. See the following subsection (\ref{sec:nltsproof}) for the proofs. They proceed similarly to that of Theorem 1 and Corollary 1 in \cite{bravyi2019obstacles} using our more general notion of symmetry.
	\newline

	It may be possible to use our Hamiltonian construction to create $k$-uniform hyperedges of $k > 3$, while keeping the \emph{partial $\mathbb{Z}_2$ symmetry}. Because of this, we expect a constant-fraction obstruction of QAOA for some instances of Max $k$XOR, at every fixed $k$, until the depth is logarithmic with $n$.
	\newline

    It is also possible that simple extensions to QAOA can avoid this obstruction, notably the so-called recursive approach (RQAOA) of \cite{bravyi2019obstacles}, given a suitable generalization of its variable elimination procedure to $k>2$.  
    We leave this as an open research direction.

\subsection{Proof of obstruction results}
\label{sec:nltsproof}
We first show the proof of Theorem~\ref{ref:nltsthmformal}. 
	Let's define the following family of Hamiltonians $H_n$. Given the subset $V_+$ with $\mathbb{Z}_2$ symmetry (such that $|V_+| = n'$), construct a sparse random $D$-regular graph $G_{n'}(V_+, E)$. We call $G_{n'}$ the ``inner graph''. We take $G$ and add $\lfloor n'/r \rfloor - O(1) \le n'/r$ nodes for some constant $r > 4$, split evenly into two sets: $A$ (the sources) and $B$ (the sinks). We replace every edge $e_{uv}$ with two hyperedges: one that includes a node $a(e_{uv}) \in A$ and one that includes a node $b(e_{uv}) \in B$. We assign hyperedges so that the maximum degree of each new node is at most $2Dr$. Then each Hamiltonian is defined as follows:
	\begin{align*}
		H_n     & = H_{n,A} - H_{n,B} + |E|                              
		        &                                                        
		H_{n,A} & = \frac{1}{2}\sum_{e_{uv}\in E} Z_{a(e_{uv})} Z_u Z_v  
		        &                                                        
		H_{n,B} & = \frac{1}{2}\sum_{e_{uv} \in E} Z_{b(e_{uv})} Z_u Z_v 
	\end{align*}
	This Hamiltonian is local, because each interaction term operates on $3 = O(1)$ qubits, and each qubit is involved in at most $2Dr = O(1)$ terms.
	\newline

	What do we know about the graphs $G_{n'}$? Since they are random graphs, they are good expanders. The Cheeger constant, a measure of a graph's connectedness, is thus lower-bounded by some $h$ for all $n'\in\mathcal{I}$, such that $(D-\lambda_2)/2 < h < \sqrt{2D(D-\lambda_2)}$ (where $\lambda_2 \ge 2\sqrt{D-1}$ is the second eigenvalue of the adjacency matrix).\footnote{This is known as Cheeger's inequality.}
	\cite{friedman2004proof} shows that random regular graphs are almost Ramanujan at large enough $D$; that is, a random regular graph has $\lambda_2 < 2\sqrt{D-1} + \delta$ for any $\delta > 0$ with probability $1-o(1)$ as $n \to \infty$.
	\newline

	Let's look at the energy of $H_n$. We know that
	\begin{align*}
		\langle x | H_n | x \rangle & = |E| + \langle x |(H_{n,A}-H_{n,B}) | x \rangle                                                                                        \\
		                            & = |E| +  \sum_{e_{uv}\in E}  \big(sgn(x_{a(e_{uv})}) - sgn(x_{b(e_{uv})})\big)  \langle x' | \frac{1}{2}  Z_u Z_v | x' \rangle          
		\\
		                            & =  \langle x' | \sum_{e_{uv} \in E} \Bigg (1 - \Big(sgn(x_{b(e_{uv})}) - sgn(x_{a(e_{uv})}) \Big) \frac{1}{2} Z_u Z_v  \Bigg )| x' \rangle 
	\end{align*}
	where $x'$ are the bits corresponding to vertices on the inner graph (i.e. not including $x_A$ and $x_B$).
	Suppose that the bit string $x$ induces an energy below $2|E|/14$. We will argue that the energy is not much lower than the energy given by $x'$, nodes in $A$ all labeled with $-1$, and nodes in $B$ all labeled with $+1$.\footnote{In fact, the only ground states of $H_n$ (i.e. with zero energy) occur when the vertices in $A$ are all $-1$, and the vertices in $B$ are all $+1$, and the vertices of the inner graph are all $+1$ or all $-1$.}
	\newline

	We can view $H_n$ as a sum of two Max 2XOR instances on the same graph. The total weight of each edge $e_{uv}$ will be $-2$, $0$, or $2$, depending on the labels of $a(e_{uv})$ and $b(e_{uv})$. The number of edges with weight $0$ limit the extremal energies; given $k$ edges with weight $0$, the minimum energy is $k$ (and maximum $2|E| - k$).
	\newline

	We assign the edges to sources and sinks in a special way.
	We choose two random orderings of edges, and assign nodes of $A$ from one ordering and nodes of $B$ from the other ordering (keeping the maximum degree below $2Dr$). This means if the energy is low enough, some labelings aren't allowed. Choose a $1/12$ fraction of vertices in $A$, and suppose their labels disagree with the rest of the labels of $A$. No matter which vertices in $A$ are chosen, the associated edges will be evenly distributed among vertices of $B$. Without flipping any labels in $B$, an average of $1/12$ fraction of edge weights will be zero.\footnote{When $2|E| = n'D$ is large enough, the fraction of zero edge weights concentrates around this value.} But flipping a $\nu$ fraction of labels in $B$ will cause even more edge weights to be zero ($1/12 + \nu(1 - 2/12) \ge 1/12$). This means the minimum energy is at least $2|E|/12$, which is not allowed by assumption. So this labeling isn't allowed. The same is true for labelings of $B$.
	\newline

	So both $A$ and $B$ have at most a $1/12$ fraction of vertices ``flipped''. How much could ``un-flipping'' the labels increase the energy? The fraction of ``off'' edges that are turning on is at most $2/12 - 2/144$, and the amount of ``reversed'' edges is at most $1/144$.  So the energy increase is at most $|E|(2/12 - 2/144) + 2|E|(2/144) = 2|E|/12$ when making all labels in $A$ the same and all labels in $B$ the same.
	\newline

	It turns out there is only one configuration that it could be close to. When the labels of $A$ and $B$ are the same sign, the energy is always $|E|$, which is not in the range. When $A$ is positive and $B$ is negative, the energy is $2|E|$ minus twice the cut of the inner graph. On a random graph, the optimal cut is less than $|E|/2 + |E|/\sqrt{D}$ at large enough $D$~\cite{dembo2017extremal}, so this is not in the range either.\footnote{We and Bravyi both use expander graphs, but our graphs are random, not bipartite.} So the only nearby configuration must be when $A$ is negative and $B$ is positive.
	\newline

	So the following is true when the energy is below $2|E|/14$:
	\begin{align*}
		\langle x | H_n | x \rangle & \ge 2\langle x' | \frac{1}{2}  \sum_{(u,v)\in E} (I - Z_u Z_v) | x' \rangle - \frac{n'D}{12} 
		\ge
		2h \min\{|x'|, n'-|x'|\} - \frac{n'D}{12}
	\end{align*}
	The last inequality comes from a fact of expander graphs in \cite{bravyi2019obstacles}.
\newline

	Assume the NLTS property is false. Then there exists a low-energy distribution where the average energy is at most $\epsilon n$. By Markov's inequality, the set of states $S_{low}$ (all bit strings with energy at most $2 \epsilon n$) have chance of occurring $p(S_{low}) > 1/2$.
	\newline

	Let $\epsilon = h/50$. Then every bit string in $S_{low}$ has energy at most $nh/25$; since $h \le D\sqrt{2}$, this energy is at most $nD\sqrt{2}/25 < n'D/14$ when $r > 4$.
	So the inequality above applies.
	Combining the equations, every bit string in $S_{low}$ satisfies
	\begin{align*}
		\min\{|x'|, n'-|x'|\} & \le \frac{n}{12} + \frac{n'D}{24h} \le \frac{n'}{9.6} + \frac{n'D}{12(D-2\sqrt{D-1}-\delta)} \le \frac{n'}{4} & \text{when $r > 4$ and $D > 20$.} 
	\end{align*}
	So the Hamming weight of the inner graph is at most $n'/4$ or at least $3n'/4$. Let's describe the sets of bit strings where the Hamming weight of the inner graph is at most $n'/4$ or at least $3n'/4$ as $S$ and $S'$, respectively. (The other $|V/V_+| \le n'/r$ nodes are free to be anything, so $dist(S, S') \ge n'/2 - 2n'/r$.) Then $S_{low} \subseteq S \cup S'$ and so $p(S_{low}) \le p(S) + p(S')$.
	\newline

	But the $\mathbb{Z}_2$ symmetry holds under the vertices of the inner graph; that is, $p(x_A,x_B,x') = p(x_A,x_B,\bar{x}')$. So there is a bijection between $x \in S$ and $x' \in S'$ such that $p(x) = p(x')$; so $p(S) = p(S')$. So $p(S) = p(S') \ge 1/4$. This combined with Corollary 43 from \cite{Eldar_2017} implies
	\begin{align*}
		0.8n(1/2 - 2/r) \le n'(1/2 - 2/r) &\le 16 n^{1/2} 8^{d/2} \\
		n & \le  400 \cdot 8^d (1/2 - 2/r)^{-2} & \text{when $r > 4$.} 
	\end{align*}
	This means the NLTS property holds whenever $n$ is larger than the value above. For example, when $r = 9$, the NLTS property holds for $n > 5184 \cdot 8^d$.


	\paragraph{Obstruction for QAOA on Max 3XOR.} We
	use Theorem~\ref{ref:nltsthmformal} to prove that sub-logarithmic depth QAOA is obstructed for a family of graphs on Max 3XOR, as stated in Corollary~\ref{ref:nltscorformal}. This
	follows similarly to the proof of Corollary 1 in \cite{bravyi2019obstacles}. Note that we have made no attempt here to optimize the constants.
	\newline

	A QAOA circuit on the Hamiltonians constructed earlier has \emph{partial $\mathbb{Z}_2$ symmetry}. Since $X^{\otimes |V_+|}$ commutes both with the mixing operator and the Hamiltonian, it commutes with the exponential of each term.
	\newline

	We can bound the circuit depth of the depth-$p$ QAOA composed of two-qubit gates as a constant times $pD$. First, note that the chromatic index of a 3-uniform graph is at most $3\Delta$, where $\Delta$ is the maximum degree of the graph \cite{OBSZARSKI201748}. We know $\Delta \le 2 r D = 18D$ when $r = 9$. And we can construct three-qubit $Z_a Z_b Z_c$ gates using 4 two-qubit gates (see Figure 6.1 of \cite{hadfield2018quantum}). So we can construct the QAOA circuit using a quantum local circuit of depth $d \le  216 pD$.
	\newline

	Now at large enough $n$, the NLTS property holds, so all inner products $\langle \psi | H | \psi \rangle > \epsilon n = hn/50$ for any $\psi$ reachable by a depth-$p$ QAOA circuit. Because $k$ is odd, this also implies $\langle \psi | H | \psi \rangle < 2|E| - \epsilon n = 2|E| - hn/50$ for any $\psi$ reachable by a depth-$p$ QAOA circuit. (If we could violate this inequality with some state $|\xi\rangle$, then $X^{\otimes n}|\xi\rangle$ would violate the first inequality.) Since $H$ corresponds to a Max 3XOR instance with $2|E| = n'D$ clauses, this bounds the maximum satisfying fraction at $1 - hn/(50n'D) < 1 - h/(50D)$.
	\newline

	Given the earlier bound on the Cheeger contant, the maximum satisfying fraction is at most
	$$
	1 - \frac{D - 2\sqrt{D-1} - \delta}{100D} = \frac{99}{100} + \frac{2\sqrt{D-1} + \delta}{100D}.
	$$
	This is a nontrivial obstruction since the instance is fully satisfiable. This occurs when $n > 5184 \cdot 8^{d}$ where $d \le 216 pD$, so it occurs for all $p < \log_2(n/5184) (648 D)^{-1}$. So a QAOA circuit with $p$ below this bound cannot achieve better than $0.99 + O(1/\sqrt{D})$ on certain fully satisfiable bounded degree Max 3XOR instances.
	
		\section{Discussion}\label{sec:conclusions}
	This paper calculates numerical upper and lower bounds of local classical and quantum algorithms approximately solving instances of Max $k$XOR. At $k > 4$ on triangle-free instances, the depth-1 QAOA is the best known local algorithm to our knowledge; however, both algorithms studied perform suboptimally on random instances at high $k$. The optimal value of a random instance is determined via its relationship to the Parisi constant of a spin glass; the numerical Parisi calculator is of independent interest. We also show that in the worst case, some low-depth quantum local circuits, including the QAOA, are obstructed on this problem for $k=3$. This is true even if the overlap gap property is not known to hold.
\newline

    This work raises many open questions, some of which are quite general. What are the best algorithms, quantum or classical, for approximately solving Max $k$XOR at higher $k$? Is there a connection between the average performance of all one-local algorithms on triangle-free instances and instances with random signs? Is there a method to compute the maximum satisfying fraction of any constraint satisfaction problem using spin glass theory?  Further study is needed here.
\newline

When $k=2$, we show an obstruction to sub-logarithmic-depth QAOA by converting obstructed Max $k$XOR instances with \emph{$\mathbb{Z}_2$ symmetry} to Max $(k+1)$XOR instances with \emph{partial $\mathbb{Z}_2$ symmetry}. It is possible this procedure works for all even $k$, and extending the proof in~\cite{bravyi2019obstacles} will lead to an obstruction for all $k \ge 2$. However, it is unknown whether any efficient algorithms, even nonlocal quantum algorithms such as RQAOA~\cite{bravyi2019obstacles}, can outperform QAOA on the obstructed instances for $k > 2$.
\newline

	Ultimately it is important to better characterize the power of quantum circuits for exact and approximate optimization. 
	A natural extension of this work is to compare depth-$p$ QAOA and depth-$p$ local classical algorithms on Max $k$XOR for $p>1$.
	However, quantum algorithms for optimization appear challenging to analyze beyond relatively simple cases. Even the QAOA proves difficult to analyze at constant-depth, let alone the more exciting $p=\Theta(\log n)$ or $p=\textrm{poly}(n)$ regimes.
	\newline

	In the meantime, the search continues for NISQ optimization advantage. Our results suggest that QAOA at depth-1 is only weakly advantageous on this problem, and that local quantum circuits are obstructed on some instances of this problem. But the QAOA or other quantum algorithms may be especially effective on certain families of Max $k$XOR instances. It remains open if there is a setting where near-term quantum computers solve an optimization problem definitively better than all classical computers; or, conversely, how much results obstructing quantum advantage can be generalized and strengthened.

	\section*{Acknowledgements}
	Luca Trevisan and Elton Yechao Zhu answered questions about classical algorithms for Max $k$XOR. Wei-Kuo Chen answered questions about the Parisi functional. Nicholas E. Sherman and Gianni Mossi gave an overview of symmetry breaking and spin glass theory. Andrea Montanari pointed us to a previous paper numerically calculating the Parisi formula. Jonathan Wurtz and Juspreet Sandhu participated in a useful discussion about the limits of local algorithms on Max $k$XOR.
	Eddie Farhi and Sam Gutmann read a draft and noted connections to instances with random signs. Casey Duckering and Konstantinos Ameranis suggested techniques to improve numerical precision. Boaz Barak read a draft and proposed a general connection between CSPs and spin glasses. K.M. thanks Eleanor Rieffel and everyone at QuAIL for listening to a preliminary talk and for a pleasant summer with the group.
\newline

    We are grateful to NASA Ames Research Center for support. K.M. participated in the Feynman Quantum Academy internship program. K.M. was supported by the National Aeronautics and Space Administration under Contract Number NNA16BD14C, managed by the Universities Space Research Association (USRA) NAMS Student R\&D Program. S.H. and K.M. were supported by the DARPA ONISQ program (DARPA-NASA IAA 8839, Annex 114).
	This material is based upon work supported by the National Science Foundation Graduate Research Fellowship Program under Grant No. DGE-1746045. Any opinions, findings, and conclusions or recommendations expressed in this material are those of the author(s) and do not necessarily reflect the views of the National Science Foundation.
	

\bibliographystyle{halpha-abbrv-mod}
\bibliography{research}
\clearpage
\newpage
	\appendix
	

	\section{
	Depth-1 QAOA for Max 3XOR}
	\label{sec:max3xorworked}
	Here we derive the performance of the depth-1 QAOA for any regular, triangle-free Max 3XOR instance; the cases of Max $k$XOR with $k>3$ are similar. 
	The $k=3$ case was first analyzed by \cite{farhi2015quantum}, who called the problem Max E3LIN2. Consider a Max 3XOR problem instance with $m$ clauses and $n$ variables. This maps to the cost Hamiltonian $C = \frac{m}{2} + \sum_{abc} d_{abc} Z_a Z_b Z_c$ where $d_{abc}$ is $0$ if there is no equation for the variables $a,b,c$, otherwise $\pm 1/2$ depending on the parity of the constraint. In a triangle-free instance, the neighbors of $Z_a$, $Z_b$, and $Z_c$ are all distinct and not any of $\{Z_a, Z_b, Z_c\}$.
	\newline

	Let's consider the 
	quantities $\langle d_{abc} Z_a Z_b Z_c \rangle$ for depth-1 QAOA. The expected satisfying fraction 
	is $1/2$ plus the average of $\langle d_{abc} Z_a Z_b Z_c \rangle$ across all clauses. Using the Pauli Solver technique from \cite{hadfield2018quantum} and \cite{Wang2018}
	we write $\langle d_{abc} Z_a Z_b Z_c \rangle = d_{abc} \Tr[\rho_0 U_C^\dagger \left(U_B^\dagger Z_a Z_b Z_c U_B \right)U_C]$ where 
	$\rho_0 = \ket{\psi_0}\bra{\psi_0}$.
	\newline

	The first step to expand the inner conjugation $U_B^\dagger Z_a Z_b Z_c U_B=e^{2i\beta(X_a + X_b + X_c)} Z_a Z_b Z_c$ using Pauli operator commutation rules. Let $p = cos(2\beta)$ and $q = sin(2\beta)$. Then using $X^2=I$ this becomes:
	\begin{align*}
		  & (pZ_a + qY_a) (pZ_b + qY_b) (pZ_c + qY_c)                                                                                              
		\\
		  & = p^3 Z_a Z_b Z_c + p^2 q (Z_a Z_b Y_c + Z_a Y_b Z_c + Y_a Z_b Z_c) + pq^2 (Z_a Y_b Y_c + Y_a Z_b Y_c + Y_a Y_b Z_c) + q^3 Y_a Y_b Y_c 
	\end{align*}
	Let's look at the outer conjugation and trace of each term separately.
	\paragraph{All $Z$s.} The term with all $Z$s will commute with $U_C$ and cancel out.
	
	\paragraph{One $Y$.} For the terms with one $Y$ (suppose $Y_a$ without loss of generality), only the $U_C$ terms involving variable $a$ do not commute. And each neighbor $Z_n$ for $n\not\in\{b,c\}$ has to be cancelled out. So each ``neighbor exponential'' contributes $cos(2d_{ak\ell} \gamma)$ for some variables $k,\ell$. Because cosine is even, and $2d_{ak\ell} = \pm 1$, this is simply $cos(\gamma)$.
	\newline

	The non-neighbor exponential ($Z_aZ_bZ_c$) will contribute:
	\begin{align*}
		  & (cos(d_{abc}\gamma) + i sin(d_{abc} \gamma)Z_a Z_b Z_c)Y_a Z_b Z_c (cos(d_{abc}\gamma) - i sin(d_{abc} \gamma)Z_a Z_b Z_c) 
		\\
		  & = 2isin(d_{abc} \gamma) cos(d_{abc} \gamma) Z_a Y_a = sin(2d_{abc} \gamma) X_a = 2d_{abc} sin(\gamma) X_a                  
	\end{align*}
	Multiplying by $d_{abc}$ causes the constant to be $1/2$. Let $s = sin(\gamma)$ and $c = cos(\gamma)$. So the terms with one $Y$ will contribute $0.5s(c^{D_a} + c^{D_b} + c^{D_c})$, where $D_i$ is the number of ``other neighbors'' of variable $i$ (i.e. the degree of variable $i$ is $D_i + 1$).
	
	\paragraph{Two $Y$s.}Consider a term with two $Y$'s (suppose $Y_a$ and $Y_b$). In a triangle-free instance, the neighbor exponentials will leave the $Y$s as $Y$s, since they need to be cancelled out. But then the $ZZZ$ exponential will always leave at least one $Z$; either $Z_c$ or $Z_aZ_b$. So this term actually has no contribution at depth-1!
	
	\paragraph{All $Y$s.} Consider the term with three $Y$'s. Here, all neighbor exponentials will contribute $c$. The $Z_aZ_bZ_c$ term will contribute:
	\begin{align*}
		  & (cos(d_{abc}\gamma) + i sin(d_{abc} \gamma)Z_a Z_b Z_c)Y_a Y_b Y_c (cos(d_{abc}\gamma) - i sin(d_{abc} \gamma)Z_a Z_b Z_c)                  
		\\
		  & = 2isin(d_{abc} \gamma) cos(d_{abc} \gamma) Z_a Y_a Z_b Y_b Z_c Y_c = -sin(2d_{abc} \gamma) X_a X_b X_c = -2d_{abc} sin(\gamma) X_a X_b X_c 
	\end{align*}
	Multiplying by $d_{abc}$ causes the constant to be $-1/2$. So this term will contribute $-0.5sc^{D_a + D_b + D_c}$.
	\newline

	All together now:
	\begin{align*}
		\langle d_{abc} Z_a Z_b Z_c \rangle & = p^2 q(0.5 s)(c^{d_a} + c^{d_b} + c^{d_c}) + q^3 (-0.5 s) c^{d_a + d_b + d_c} 
	\end{align*}
	\paragraph{Regular graphs.}
	Let's assume we're working with $(D+1)$-regular graphs. Then all $D_a = D_b = D_c = D$ for all subgraphs. So the satisfying fraction is $1/2 + \langle d_{abc} Z_a Z_b Z_c \rangle$. We can find the optimal value of the depth-1 QAOA by finding where $d/d\beta = 0$:
	\begin{align*}
		\langle d_{abc} Z_a Z_b Z_c \rangle & = 0.5 qs(3p^2 c^D - q^2  c^{3D})   
		\\
		                                    & = 0.5s c^D(3q(1-q^2) - q^3 c^{2D}) 
		\\
		                                    & = 0.5 sc^D (3q - q^3(3 + c^{2D}))  
	\end{align*}
	\begin{align*}
		d/d\beta = 0 & \xrightarrow[]{} (dq/d\beta) (3 - 9q^2 - 3q^2 c^{2D}) = 0 
		\\
		q^2          & = \frac{3}{3c^{2D} + 9} = \frac{1}{c^{2D} + 3}            
	\end{align*}
	Since $c^{2D} = (c^D)^2 \ge 0$, $q$ will always be real. We use this to identify the optimal value:
	\begin{align*}
		\langle d_{abc} Z_a Z_b Z_c \rangle_{max} = 0.5 sc^D (2q) = \frac{sc^D}{\sqrt{c^{2D}+3}} = \frac{s}{\sqrt{1 + 3/c^{2D}}} = \frac{s}{\sqrt{1 + 3/(1-s^2)^D}} 
	\end{align*}
	The maximum with respect to $\gamma$ can be found by setting $d/d\gamma = 0$:
	\begin{align*}
		(s'/s)f + f(-1/2)(1 + 3/(1-s^2)^D)^{-1}(3/(1-s^2)^{D+1})(-D)(-2s)s' = 0 
	\end{align*}
	\begin{align*}
		1/s             & = \frac{3sD/(1-s^2)^{D+1}}{1 + 3/(1-s^2)^D} 
		\\
		1 + 3/(1-s^2)^D & = 3s^D/(1-s^2)^{D+1}                        
		\\
		(1-s^2)^D       & = (3sD)/(1-s^2) - 3                         
	\end{align*}
	\begin{figure}[tbp]
		\centering
		\includegraphics[width=5in]{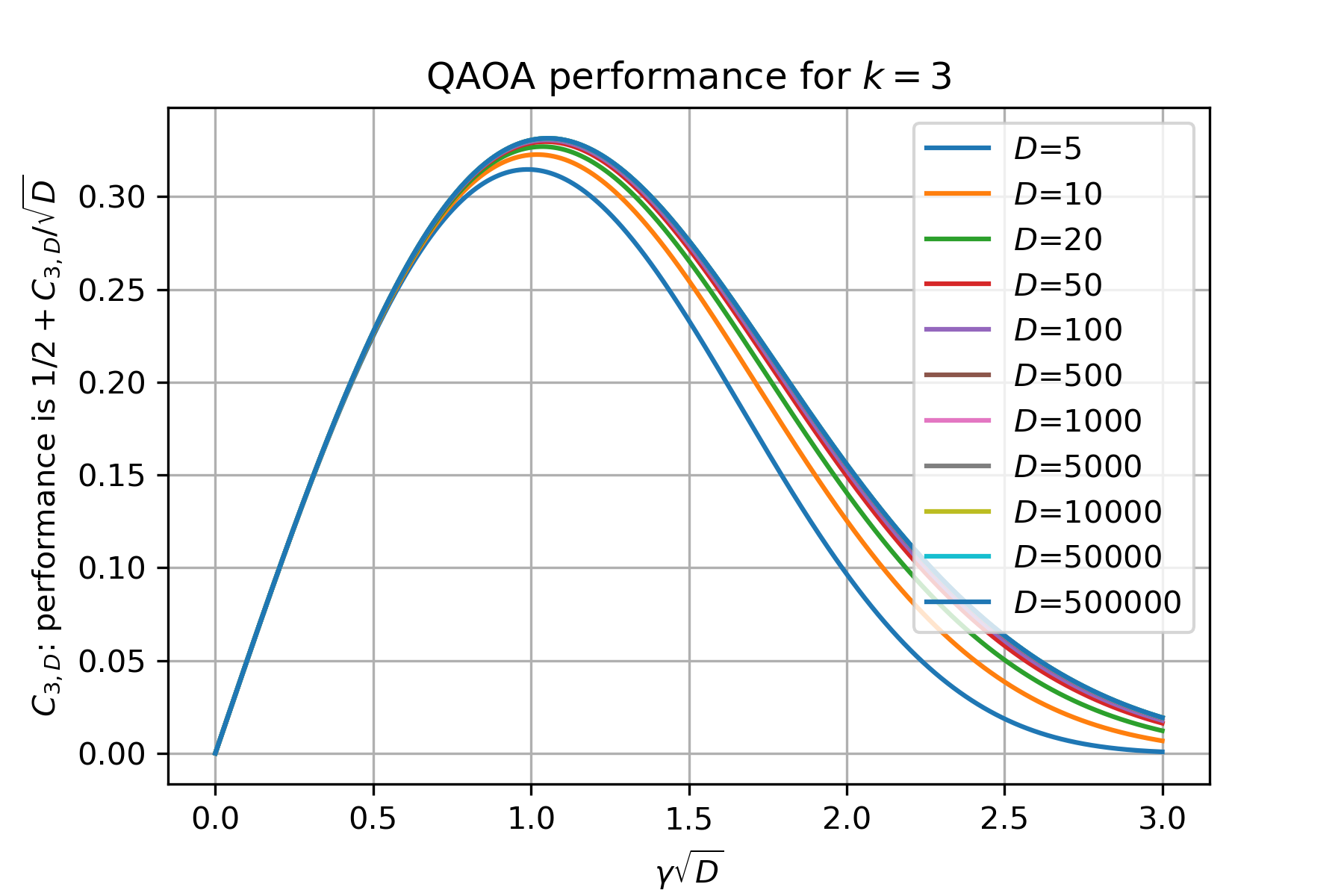}
		\caption{ Color online. The performance of the depth-1 QAOA on $(D+1)$-regular triangle-free Max 3XOR instances. The optimal constant $C_{3,D}$ increases with $D$. At its maximum, the depth-1 QAOA has average satisfying fraction approximately $1/2 + 0.33/\sqrt{D}$. The optimal $\gamma$ is proportional to $1/\sqrt{D}$.}
		\label{fig:qaoamax3xor}
	\end{figure}
	\newline
	This can be numerically solved at every $D$. See Figure \ref{fig:qaoamax3xor} for a plot of optimal performance across $\gamma$. Note that $\gamma$ tends to zero as $D$ gets large. In this regime, we can take a small angle approximation, where $s \approx \gamma = k/\sqrt{D}$:
	\begin{align*}
		\frac{k}{\sqrt{D}}\frac{1}{\sqrt{1 + 3/(1 - k^2/D)^D}}  
		\approx \frac{1}{\sqrt{D}}\frac{k}{\sqrt{1 + 3e^{k^2}}} 
	\end{align*}
	This is maximized at $e^{k^2}(1+k^2) = -1/3$. Numerically, we find optimal values $k\approx 1.0535$ and $\beta \approx 0.29$, giving asymptotic performance at $\approx 0.3314627/\sqrt{D}$ above $1/2$. This is larger than the asymptotic performance calculated in \cite{farhi2015quantum}, where they set $\beta = \pi/4$ and achieve satisfying fraction $\approx 1/2 + 0.18/\sqrt{D}$.
	
	\paragraph{Lower bound for bounded number of clauses per variable.}
	When the graph is not regular but still has at most $D+1$ clauses per variable, the performance is at least as high as when the graph is $(D+1)$-regular. Consider the average value of $d_{abc} Z_a Z_b Z_c$:
	\begin{align*}
		\langle d_{abc} Z_a Z_b Z_c \rangle & = 0.5qs(p^2(c^{D_a} + c^{D_b} + c^{D_c}) - q^2(c^{D_a+D_b+D_c}))                                          
		\\
		                                    & = 0.5s((q-q^3)(c^{D_a} + c^{D_b} + c^{D_c}) - q^3(c^{D_a+D_b+D_c}))                                       
		\\
		                                    & = 0.5sq((c^{D_a} + c^{D_b} + c^{D_c}) - \frac{c^{D_a} + c^{D_b} + c^{D_c} + c^{D_a+D_b+D_c}}{c^{2D} + 3}) 
		\\
		                                    & = 0.5sq((c^{D_a} + c^{D_b} + c^{D_c})\frac{c^{2D} + 2}{c^{2D} + 3} - \frac{c^{D_a+D_b+D_c}}{c^{2D} + 3})  
	\end{align*}
	Let's look at the term with $D_a$ in it. This is $c^{D_a}$, multiplied by the following factor:
	\begin{align*}
		\frac{sq(c^{2D} + 2 - c^{D_b+D_c})}{c^{2D} + 3} 
	\end{align*}
	As we saw, the optimal $\gamma \in (0, \pi/2)$, so $s$ and $c$ are both positive, and $\beta \in (0,\pi)$ so $q$ is positive. Then this factor is nonnegative. This means we can multiply it by $c = cos(\gamma)$ until $c^{D_a}$ becomes $c^D$, this term's value will not increase! The same will be true for the other terms $D_b$ and $D_c$. So the average value of this expression is at least as high as the average value found in the $(D+1)$-regular case. This means the performance of the depth-1 QAOA on $(D+1)$-regular instances is a lower bound for the performance on the bounded constraint problem.

\clearpage
\newpage
	\section{The Parisi value for Max $k$SAT}
	\label{sec:ksat}
	There is a different set of Parisi values $B(k)$ for random $k$SAT when the formula is unsatisfiable, as shown by \cite{panchenko2017ksat}.
	When $k=3$, we get an optimal satisfying fraction of approximately $7/8 + 0.277/\sqrt{a}$, where $M=aN$ clauses. We evaluated $B(k)$ for $k \in \{3,...,9\}$, and the associated values are in  Table \ref{tab:ksat}. At higher $k$ we experienced convergence issues. The code is \href{https://nbviewer.jupyter.org/github/marwahaha/QuAIL-2021/blob/main/parisi.ipynb}{online at this link}.
	
\begin{table}[htbp]
\centering
\begin{tabular}{|c|c|c|}
\hline
$k$ & $B(k)$             & $C_k$                  \\ \hline
3   & 2.2176 & 0.277   \\ \hline
4   & 3.7457  & 0.234  \\ \hline
5   & 5.8483  & 0.182   \\ \hline
6   & 8.7320  & 0.136   \\ \hline
7   & 13.239 & 0.103  \\ \hline
8   & 18.362 & 0.071  \\ \hline
9   & 26.246 & 0.051  \\ \hline
\end{tabular}
\caption{ Calculated Parisi values $B(k)$ for the mixed-spin glass describing Max $k$SAT. The optimal satisfying fraction of a sparse random instance is $1 - 1/2^k + C_k/\sqrt{\alpha}$, given $n$ variables and $m = \alpha n$ clauses. All values are truncated. The Parisi value $B(k)$ is $C_k 2^k$.}
\label{tab:ksat}
\end{table}
\end{document}